\documentclass{article}
\usepackage[utf8]{inputenc}
\usepackage{times}
\usepackage{amsfonts}
\usepackage{amsmath,amssymb}
\usepackage{indentfirst}
\usepackage{wrapfig}
\usepackage{graphicx}
\usepackage{graphics,color}
\usepackage{caption}
\usepackage[toc,page]{appendix}
\usepackage{color}
\usepackage{cite}
\usepackage[inactive]{srcltx}
\usepackage[T1]{fontenc}
\usepackage{float}
\usepackage{hyperref}
\usepackage{subfigure}

\begin{document}
\date{}
\begin{center}
{\Large\bf Two coupled qubits under the influence of a minimal, phase-sensitive environment}
\end{center}
\begin{center}
{\normalsize G.L. De\c cordi and A. Vidiella-Barranco \footnote{vidiella@ifi.unicamp.br}}
\end{center}
\begin{center}
	{\normalsize{$^{a}$School of Electrical and Computer Engineering, University of Campinas}}\\
	{\normalsize{ 13083-852   Campinas,  SP,  Brazil}}\\
	{\normalsize{$^{b}$Gleb Wataghin Institute of Physics - University of Campinas}}\\
	{\normalsize{ 13083-859   Campinas,  SP,  Brazil}}\\
\end{center}
\begin{abstract}
In this work, we investigate the influence of a minimal, phase-sensitive environment on a system of two coupled qubits. 
The environment is constituted by a single-mode field initially prepared in a type of Schr\"odinger cat state, a quantum 
superposition of two squeezed coherent states. We present an analytical solution to the model and investigate the degradation 
of the quantum features of the system due to the action of the environment. In particular, we find that the time-averaged linear 
entropy for long times, $\bar{S}_T$,  has approximately a linear dependence on Mandel's $Q$ parameter as well as on the variance of the $\hat{X}$ quadrature of the initial state of the environment.
\end{abstract}
\section{Introduction}
\label{sec:section1}

The development of quantum technologies \cite{barnett17} demands a high level of control. However, physical systems 
are normally subjected to the action of external environments which can disrupt the coherent evolution needed, e.g., for performing 
certain quantum tasks. We thus need an accurate description of the system-environment interactions in order to mitigate the possible
destructive environmental effects such as decoherence \cite{giulini03}. Environments are normally modelled by a thermal reservoir, 
namely, a large number of quantum subsystems that are assumed to be coupled to the system of interest, normally giving rise to decoherence. 
Nonetheless, physical systems may also be subjected to environments having a small number of degrees of freedom (small environments), and remarkably, 
an environment just of the size of the system is enough to cause decoherence \cite{eisert18}, challenging the view that baths should
necessarily have a large dimension in order to do so. In fact, it has been experimentally demonstrated that a single electron can constitute a
minimal environment causing decoherence in a $\mbox{H}_2$ molecule system \cite{dorner07}. There are discussions in the literature about the 
disturbances suffered by quantum systems due to their interaction with small, uncontrollable environments constituted by either a few two-level 
systems \cite{roversi03,palma08,lombardi10,vidiella14,decordi18,mirkin21} or a few modes of the electromagnetic field 
\cite{aguiar05,talkner09,ashhab14,vidiella16,decordi20}. Also, the contribution of a single two-level system to the decoherence of a qubit 
in combination with a thermal bath has been addressed in \cite{ashhab06a,ashhab06b}. Consider a simple quantum model of two interacting 
systems, namely two dipole-dipole coupled qubits (2-level atoms) under the influence of an external environment. This system was recently 
investigated in detail considering thermal baths constituted either by a single mode field \cite{decordi20} (minimal environment) or a multimode
field \cite{decordi17} (large environment). Interestingly, we have shown that even in the former case \cite{decordi20} it can happen
the disappearance of quantum entanglement in a finite time, the Entanglement Sudden Death \cite{horodecki01,eberly04}. This is sometimes 
followed by a revival of entanglement, the Entanglement Sudden Birth. In addition to that, other quantum properties of the system such as its 
state purity and quantum coherence can be substantially affected by small thermal (phase-insensitive) environments.

In this work we are going to discuss some dynamical features of an analytically solvable model of two coupled qubits under the influence of a 
phase-sensitive environment constituted by a single mode field. We are going to consider only one of the qubits (qubit 2) coupled to the field,
keeping the qubit 1 ``isolated". The field (environment) will be assumed to be prepared in a kind of Schr\"odinger cat state \cite{dodonov74,vidiella92}, 
a quantum superposition of two squeezed coherent states, namely, $\left|\alpha,\,\xi\right\rangle$ and $\left|-\alpha,\,\xi\right\rangle$, where $\alpha$ 
is the coherent amplitude, and $\xi = r e^{i\theta}$ it is the squeezing parameter. It is well known that a variety of quantum states of 
light can be approximated by superpositions of coherent (or squeezed) states \cite{domokos94}, and we can mention a work in which the dynamics of a qubit 
coupled to a (multimode) bath initially prepared in a Schr\"odinger cat state has been investigated in \cite{hanggi10}. The single-mode environment we are 
going to consider here has additional flexibility, in the sense that it is characterized not only by the mean photon number of the initial field, 
$\langle\hat{n}\rangle$, and a relative phase $\varphi$, but also by the phase of the squeezed state $\theta$ and $c$ $(\sqrt{1 - c^2})$, the relative weights 
in the superposition state. This will allow us to model different types of small environments having distinct statistical properties. 
Our aim here is to investigate how the dynamics of the two-qubit system is affected when those parameters are changed while keeping both 
$\langle\hat{n}\rangle$ and the system-environment coupling strength $g$ fixed. In particular, we are interested in analyzing in which ways the environment 
degrades important nonclassical properties of the system of qubits, for instance, i) the state purity of the ``isolated" qubit (qubit 1), 
ii) the quantum coherence of the two-qubit state, and iii) the quantum entanglement between the two qubits. We will show that 
the statistical properties of the initial state of the environment, e.g., the fluctuations in photon number and quadratures 
(which in turn depend on $\theta$) have a significant impact on the degradation of the quantum properties of the two-qubit system. 

Our paper is organized as follows: in Section (2) we obtain the analytical solution for the model of two coupled qubits interacting with
a single mode of the field prepared in a quantum superposition of squeezed states. In Section (3) we investigate how the evolutions
of the linear entropy of qubit 1, the quantum coherence and entanglement of the two qubits are influenced if one considers different
parameters of the environment, namely, the phase $\theta$ of the squeezed coherent state and the relative weights in the superposition.
We present our conclusions in Section (4), and details of the calculations are shown in the Appendix.



\section{The system model}
\label{section01}
Our model consists of two coupled 2-level systems (qubit 1 and qubit 2) having one of them (qubit 2) interacting with a 
single-mode quantized electromagnetic field. The Hamiltonian $H$ of the system in units of $\hbar$ can be written as \cite{decordi20}
\begin{equation}\label{hamiltot}
	H=H_{0}+H_{1},
\end{equation}
where the part corresponding to the qubits' excitations plus the free field term is
\begin{equation}\label{hamilqubits}
	H_{0}=\frac{\omega}{2}\sigma_{1z}+\frac{\omega}{2}\sigma_{2z}+\omega a^\dagger a ,  
\end{equation}
and 
\begin{equation}\label{hamilqubitfield}
	H_{1}=\lambda\left(\sigma_{1}^{+}\sigma_{2}^{-}+\sigma_{1}^{-}\sigma_{2}^{+}\right)+g\left(a\sigma_{2}^{+}+a^\dagger\sigma_{2}^{-}\right)
\end{equation}
is the interaction part, under the rotating-wave approximation. The first term, with coupling strength $\lambda$, corresponds to the dipole-dipole 
coupled qubits, whereas the second term, coupling strength $g$, refers to the interaction of qubit 2 with the field, basically a Jaynes-Cummings Hamiltonian. 
We assume for simplicity the transition frequencies of the qubits being the same and in resonance with the field, i.e., 
$\omega_1 = \omega_2 = \omega_f \equiv \omega$.
The populations, as well as the qubits raising and lowering operators, are respectively given by
$\sigma_{iz}=\left|e_{i}\right\rangle \left\langle e_{i}\right|-\left|g_{i}\right\rangle \left\langle g_{i}\right|$,
$\sigma_{i}^{+}=\left|e_{i}\right\rangle \left\langle g_{i}\right|$ ($i=1,2$), and $\sigma_{i}^{-}=\left|g_{i}\right\rangle \left\langle e_{i}\right|$, 
whereas $a$ and $a^{\dagger}$ are the field's usual annihilation and creation operators.

We consider an initial separable state $|\Psi(0)\rangle = |\psi_{q1}(0)\rangle \otimes |\psi_{q2}(0)\rangle \otimes |\psi_{f}(0)\rangle$, 
with qubit 1 in its excited state, $\left|\psi_{q1}\left(0\right)\right\rangle =\left|e_{1}\right\rangle $, qubit 2 in its ground state, 
$\left|\psi_{q2}\left(0\right)\right\rangle =\left|g_{2}\right\rangle $, and the field in a quantum superposition 
of squeezed coherent states of the form
\begin{equation}\label{initialfield}
	\left|\psi_{f}\left(0\right)\right\rangle =\mathcal{N}\left[c\left|\alpha,\,\xi\right\rangle +e^{i\,\varphi}\sqrt{1-c^{2}}\left|-\alpha,\,\xi\right\rangle \right], 
\end{equation}
with $\mathcal{N}=\left[1+2c\sqrt{1-c^{2}}\cos\left(\varphi\right)\exp\left[-2\left|\alpha\mu+\alpha^{*}\nu\right|^{2}\right]\right]^{-1/2}$.
Here $\mu =\cosh r$ and $\nu=e^{i\,\theta}\sinh\left(r\right)$, being $\xi = r e^{i\theta}$ the squeezing parameter, and $\varphi$ is a relative
phase in the superposition. The squeezed coherent states are defined as:
\begin{equation}
	\left|\pm\alpha,\,\xi\right\rangle =D\left(\pm\alpha\right)S\left(\xi\right)\left|0\right\rangle,  
\end{equation}
being $D\left(\alpha\right)=\exp(\,\alpha\,a^{\dagger}-\,\alpha^{*}a)$ and 
$S\left(\xi\right)=\exp\left[\frac{1}{2}\left(\xi^{*}a^{2}-\,\xi\,a^{\dagger2}\right)\right]$ 
the displacement and squeezing operators, respectively. A scheme of generation of a superposition of squeezed coherent states 
of that type in a cavity has been presented in \cite{almeida10}. We are assuming the specific state in Eq.(\ref{initialfield}) as the initial
state of the environment bearing in mind that it can approximate a variety of states of light, depending on the choice of parameters \cite{domokos94}, making
possible to mimic the influence of environments having distinct statistical properties. Thus, our choice for the initial state $\left|\psi_{f}\left(0\right)\right\rangle$ 
is convenient for modelling different types of environments. The reader can find in the Appendix details regarding the analytical solution of model, i.e., 
the calculation of the time-dependent state vector of the system $+$ environment (two qubits $+$ field) in the interaction representation, $|\Psi(t)\rangle_I$.

\section{Influence of a small phase-sensitive environment}
\label{section02}

Now we would like to study the influence of the environment (single-mode light field) on typical quantum properties of the two-qubit system. We will
analyze the temporal evolution of the state purity of qubit 1 (linear entropy), as well as the quantum
entanglement (concurrence) and quantum coherence ($l_1$-norm of coherence) of the two-qubit system.

\subsection{Evolution of the qubit 1 linear entropy}
\label{section021}

Firstly we will focus on the dynamics of the state purity of qubit 1, quantified by the linear entropy 
$S\left(t\right)=1-Tr_{q1}\left(\rho_{q1}^{2}\right)$, which is actually a measure of mixedness of quantum states. 
Here $\rho_{q1}(t) = \mbox{Tr}_{q2,f}\left[|\Psi (t)\rangle_{I}{}_{I}\langle \Psi (t) |\right]$ 
is the reduced density operator of qubit 1 obtained by tracing both the field (environment) and qubit 2 variables. For any pure state $S = 0$,
whereas if $S > 0$, the quantum state of the system is a statistical mixture of pure states (mixed state).
In our model the linear entropy can be written as
\begin{equation}
	S\left(t\right) =\frac{1}{2}-\frac{W\left(t\right)^{2}}{2}-2\left|\rho_{eg}\left(t\right)\right|^{2},
\end{equation}
where the atomic inversion is $W\left(t\right)=1-2\rho_{gg}\left(t\right)$. The expressions for the matrix elements 
$\rho_{gg}$ and $\rho_{eg}$ can be found in the Appendix.

\begin{figure}[htpb]
	\centering
	\subfigure{\includegraphics[scale=0.40]{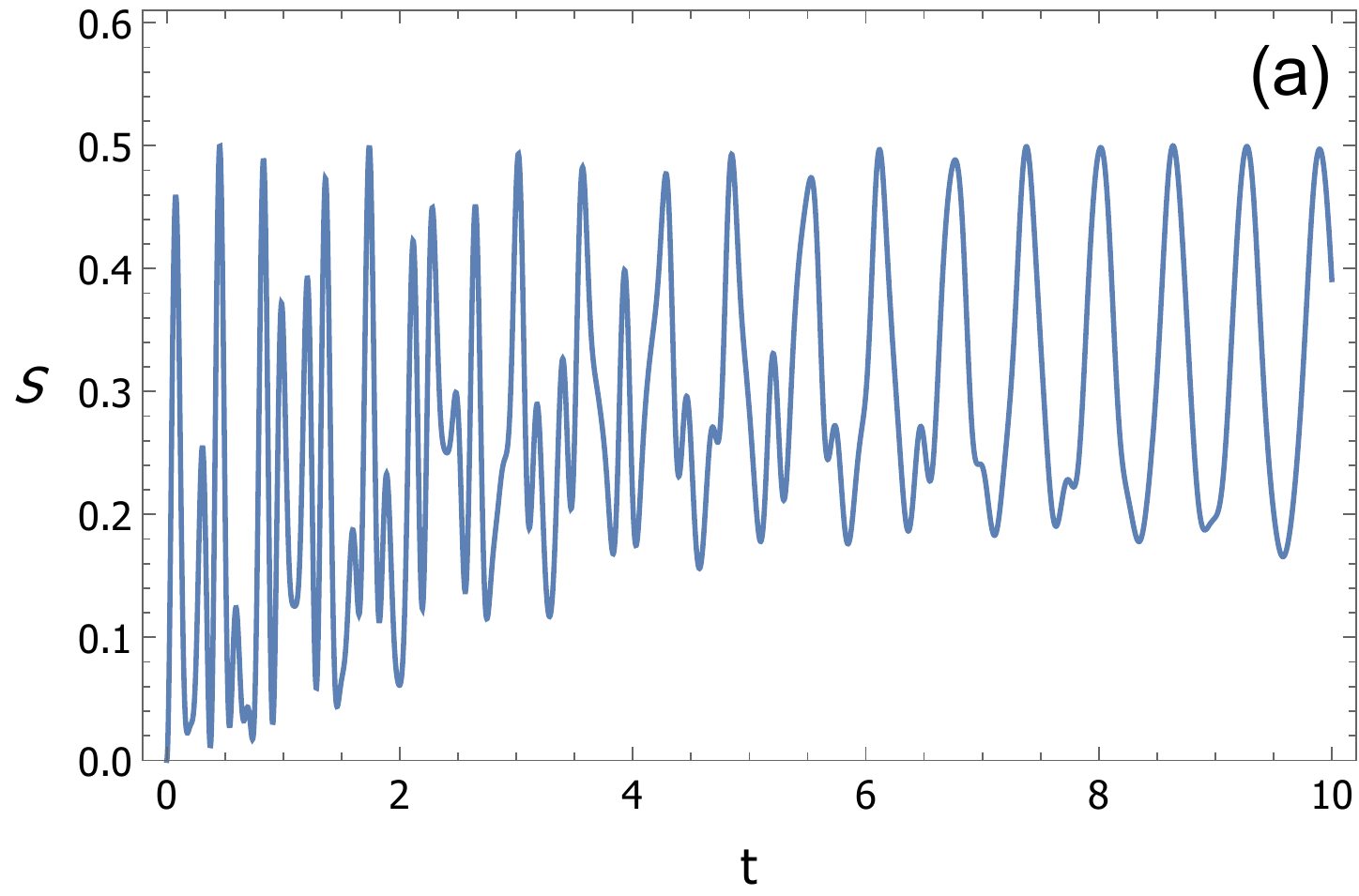}}\qquad\qquad
	\subfigure{\includegraphics[scale=0.40]{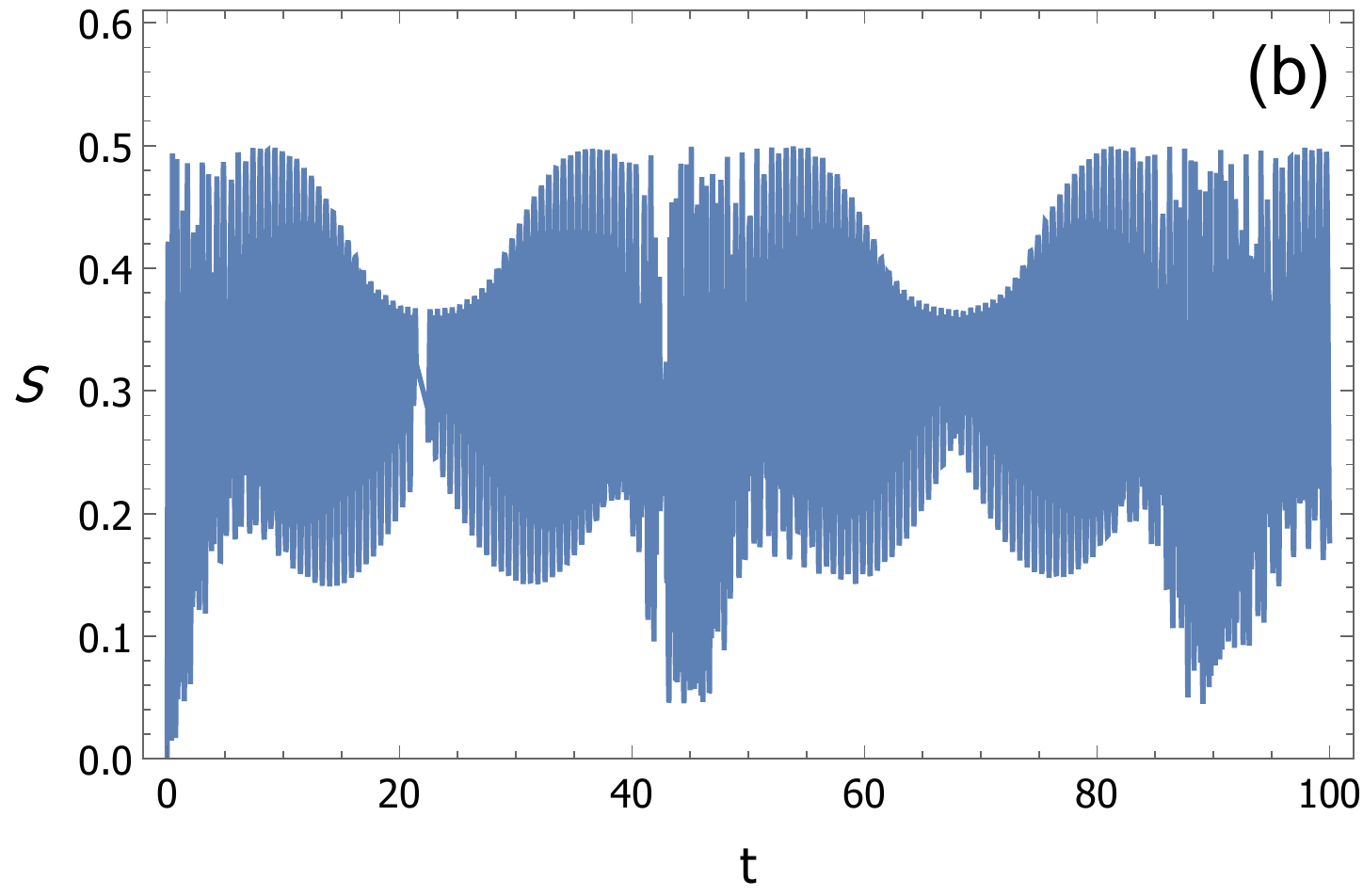}}\qquad
	\caption{Linear entropy of qubit 1 as a function of time for $c = 0$ (single squeezed state) and $\theta = 0$ (amplitude squeezing) 
		on a short time-scale (a), and a long time-scale (b). Here, $r = 1.0$, $\alpha = 5.0$ and $\varphi = \pi$.}
	\label{linearentropy01}
\end{figure}

\begin{figure}[htpb]
	\centering
	\subfigure{\includegraphics[scale=0.40]{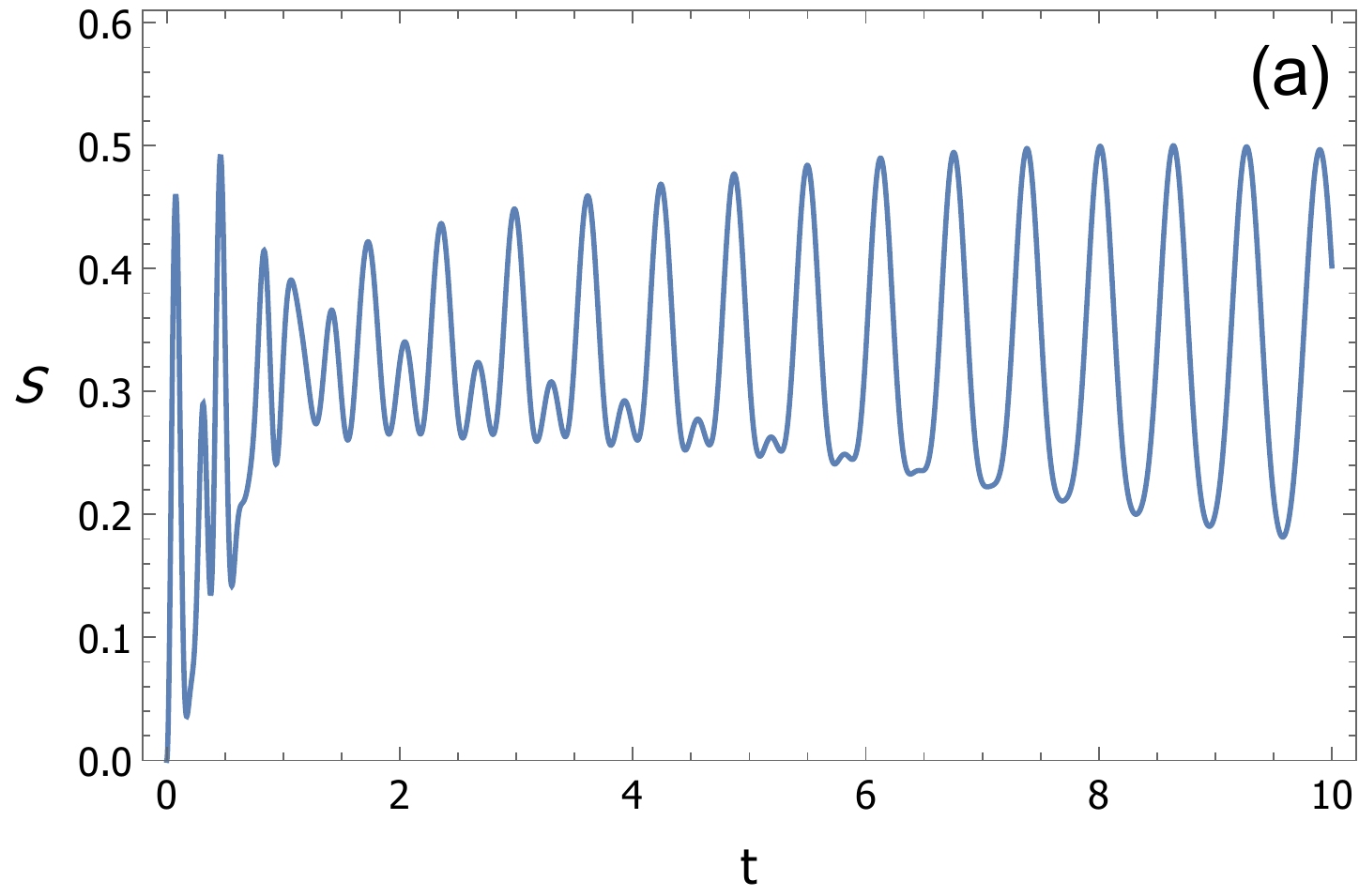}}\qquad\qquad
	\subfigure{\includegraphics[scale=0.40]{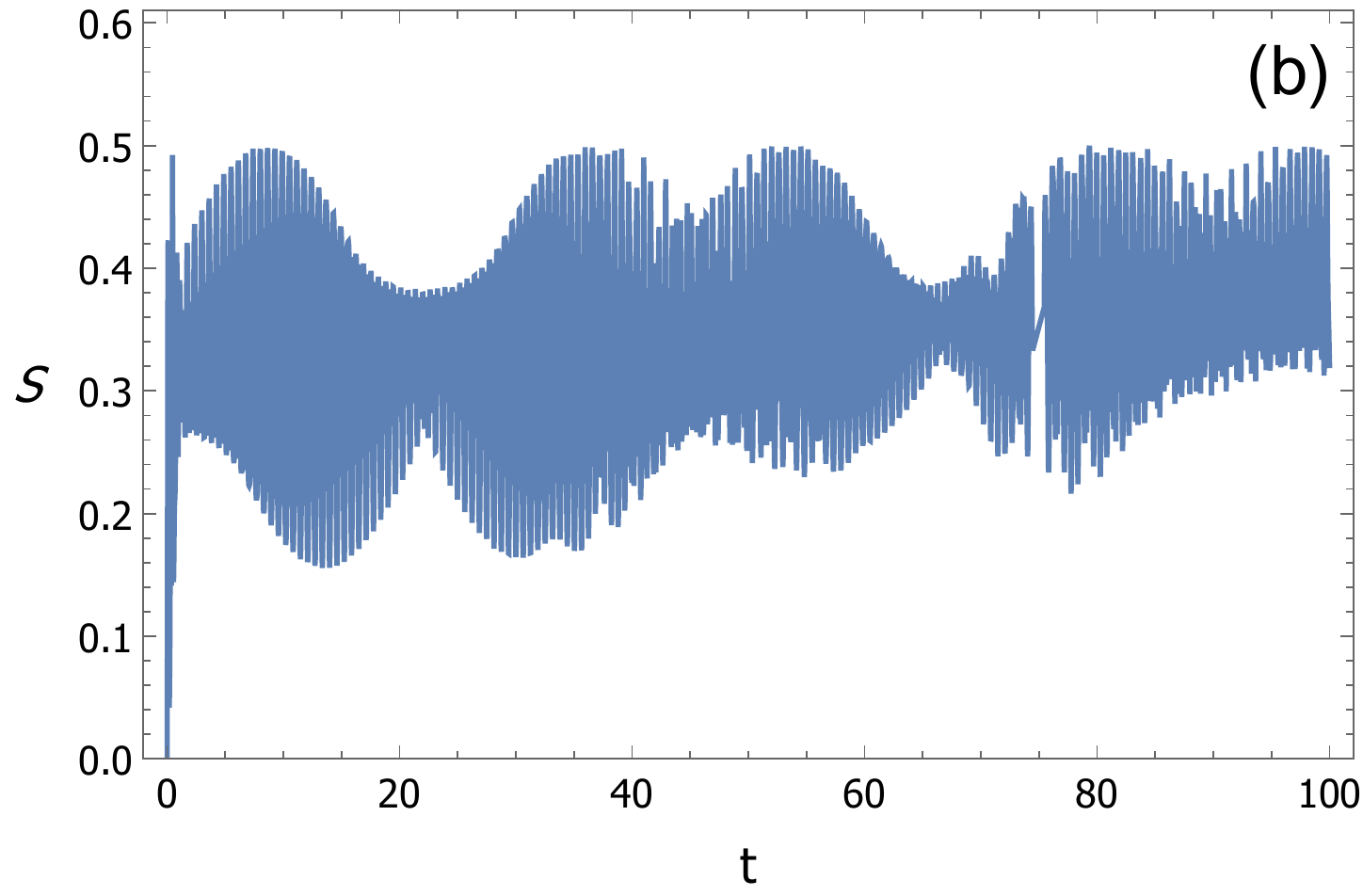}}\qquad
	\caption{Linear entropy of qubit 1 as a function of time for $c = 0$ (single squeezed state) and $\theta = \pi$ (phase squeezing) 
		on a short time-scale (a), and a long time-scale (b). Here, $r = 1.0$, $\alpha = 5.0$ and $\varphi = \pi$.}
	\label{linearentropy02}
\end{figure}

\begin{figure}[htpb]
	\centering
	\subfigure{\includegraphics[scale=0.40]{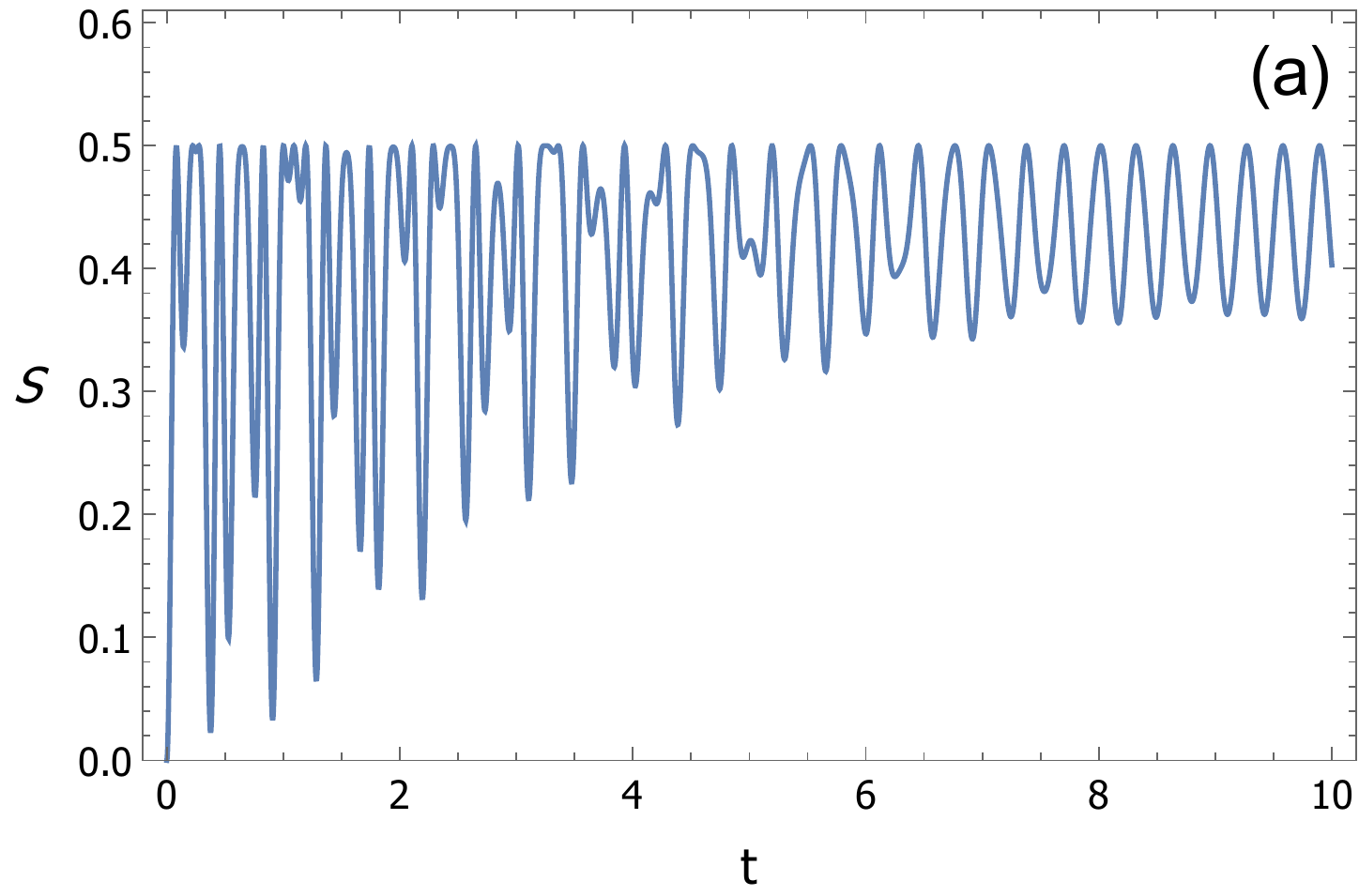}}\qquad\qquad
	\subfigure{\includegraphics[scale=0.40]{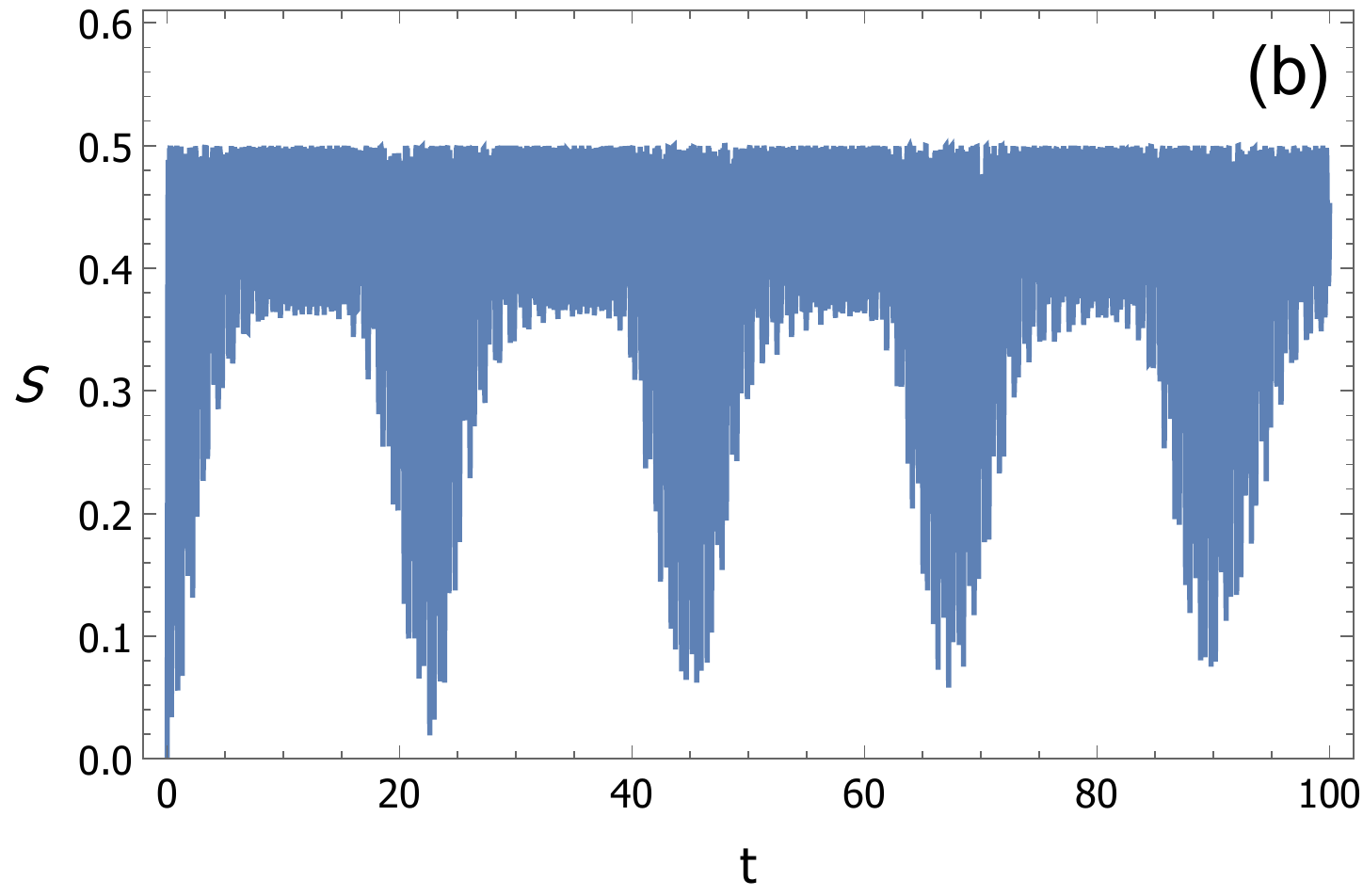}}\qquad
	\caption{Linear entropy of qubit 1 as a function of time for $c = 1/\sqrt{2}$ (equally weighted superposition) and $\theta = 0$ 
		(amplitude squeezing)  on a short time-scale (a), and a long time-scale (b). Here, $r = 1.0$, 
		$\alpha = 5.0$ and $\varphi = \pi$.}
	\label{linearentropy03}
\end{figure}

\begin{figure}[htpb]
	\centering
	\subfigure{\includegraphics[scale=0.40]{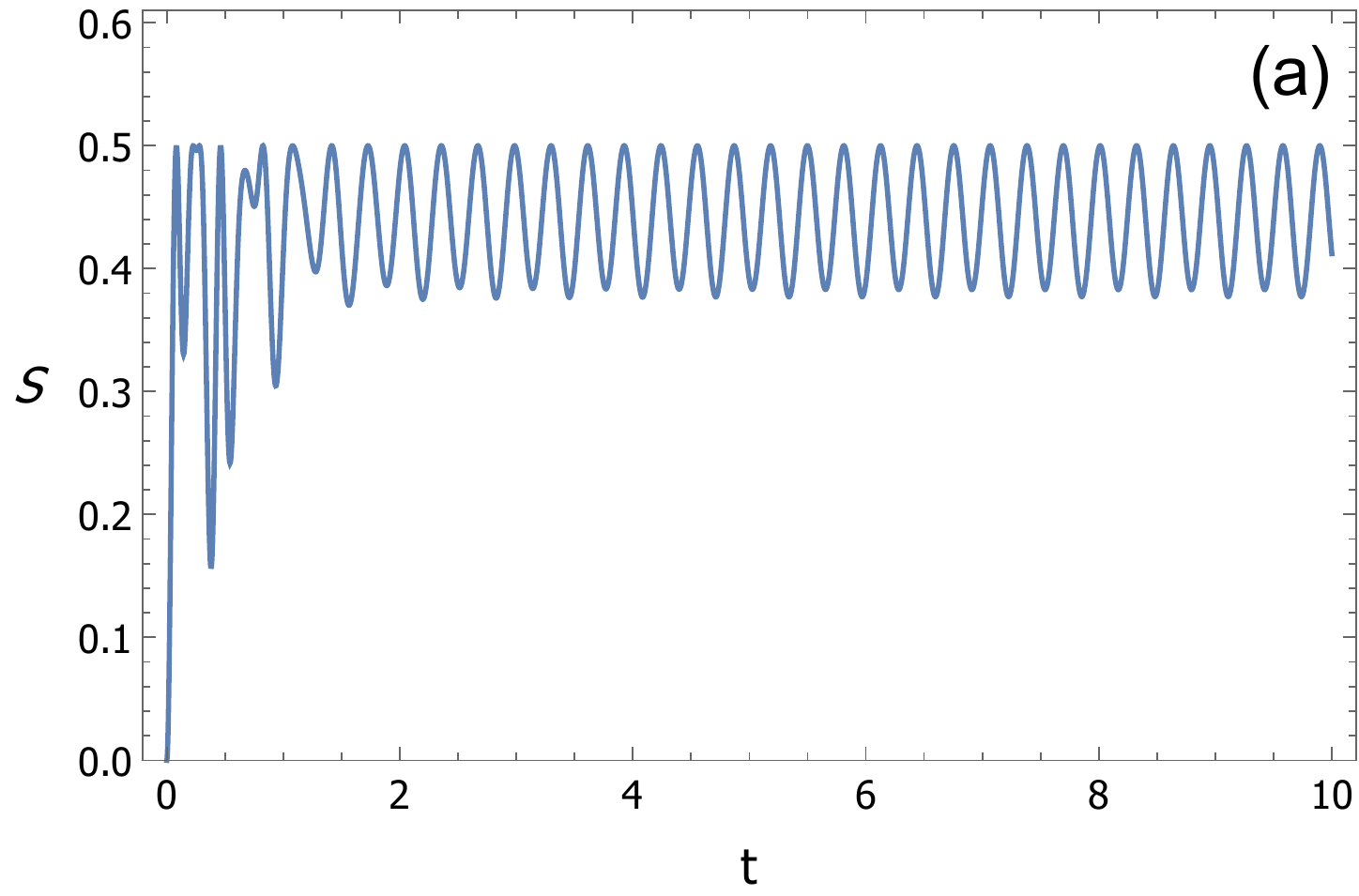}}\qquad\qquad
	\subfigure{\includegraphics[scale=0.40]{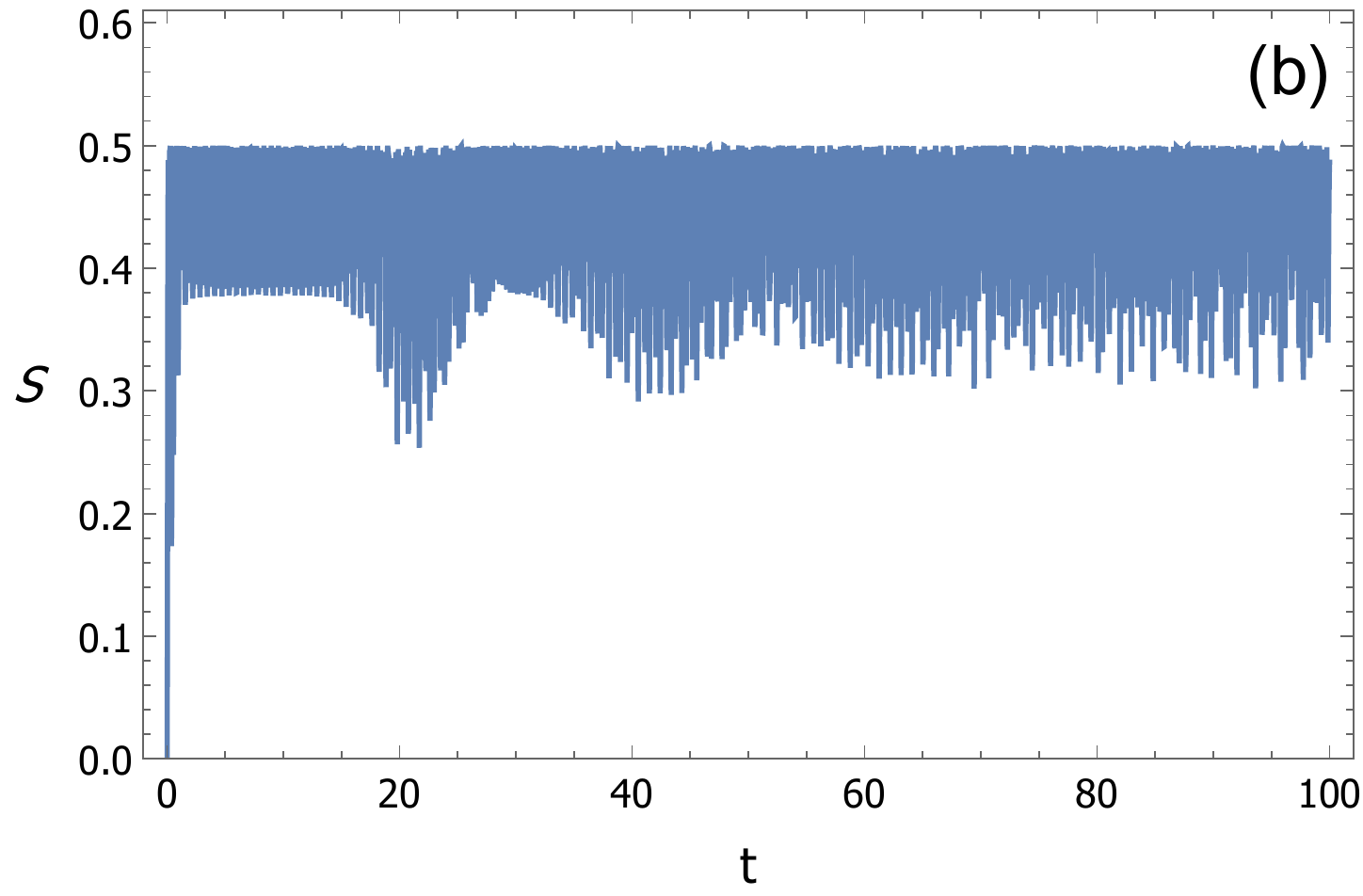}}\qquad
	\caption{Linear entropy of qubit 1 as a function of time for $c = 1/\sqrt{2}$ (equally weighted superposition) and $\theta = \pi$ 
		(phase squeezing) on a short time-scale (a), and a long time-scale (b). Here, $r = 1.0$, 
		$\alpha = 5.0$ and $\varphi = \pi$.}
	\label{linearentropy04}
\end{figure}

We make a numerical analysis of the linear entropy as a function of time, $S(t)$, 
considering that the environment has a fixed value of its initial mean excitation number, 
$\langle\hat{n}\rangle \equiv \langle\psi_{f}(0)|\hat{a}^\dagger\hat{a}|\psi_{f}(0)\rangle$. For the state in 
Eq.(\ref{initialfield}), we have that
\begin{eqnarray}
	\langle\hat{n}\rangle &=& \mathcal{N}^{2}\{ \left|\alpha\right|^{2}+\sinh^{2}r
	+2c \sqrt{1-c^2} \,e^{-2\left|\alpha\right|^{2}\left[\cosh 2r+\cos\theta \sinh 2r\right]} \\ \nonumber
	&\times& [ \sinh^{2} r -\left|\alpha\right|^{2}\cosh 4r - \left|\alpha\right|^{2}\cos\theta \sinh 4r]
	\cos\varphi\}, 
\end{eqnarray}
with $\mathcal{N}$ being a normalization constant. Note that $\langle\hat{n}\rangle$ depends on various parameters: 
the magnitudes of the coherent amplitude $\left|\alpha\right|$ and the squeezing parameter $r$, the relative phase $\varphi$, 
the phase $\theta$ and the relative weight $c$. In order to simplify our analysis, we fixed the values of some quantities in all 
calculations: $\alpha = 5.0$ (real), $r = 1.0$ and  $\varphi = \pi$, which gives us $\langle\hat{n}\rangle = 26.38$. However, 
we used different values of $\theta$ and  $c$ in order to study in which way these two quantities could influence the behaviour of 
the system. We know that in the absence of an environment (two isolated coupled qubits), $S(t)$ is a periodic function of time, given that 
the qubits get entangled/disentangled periodically as time goes on. Yet, we expect deviations from that behaviour due to the coupling 
to the minimal environment. In Fig.(\ref{linearentropy01}) we have plotted the linear entropy as a function of time for $c = 0$ and the 
squeezed state phase $\theta = 0$ for both short and long time-scales. The linear entropy of the 
qubit 1 state, initially pure $(S(0) = 0)$, displays relatively large amplitude oscillations as a function of time, meaning that
the quantum state of qubit 1 remains basically mixed. However, the oscillatory pattern is modified if we have an initial phase squeezed state 
($\theta = \pi$), instead. As shown in Fig.(\ref{linearentropy02}), there is a clear tendency for the qubit 1 state to become more mixed 
in average, during the evolution. 
Noticeable changes also occur if a second squeezed coherent state comes into play. In Fig.(\ref{linearentropy03}) we have plotted the linear entropy as a 
function of time for $c = 1/\sqrt{2}$, i.e., for an environment constituted by an equally weighted superposition of squeezed coherent states. 
A different initial state has an evident effect in the oscillatory pattern of the linear entropy, as we see in the plots. For $\theta = 0$, the 
oscillations can reach an appreciable amplitude, while if $\theta = \pi$ they are restricted to a smaller range of values, as shown
in Fig.(\ref{linearentropy04}). Thus, in addition to the influence of the squeezed state phase $\theta$, there is a significant impact
on the linear entropy dynamics if the environment is initially prepared in a superposition of squeezed coherent states (squeezed Schr\"odinger
cat state). 

\begin{figure}[htpb]
	\centering
	\subfigure{\includegraphics[scale=0.40]{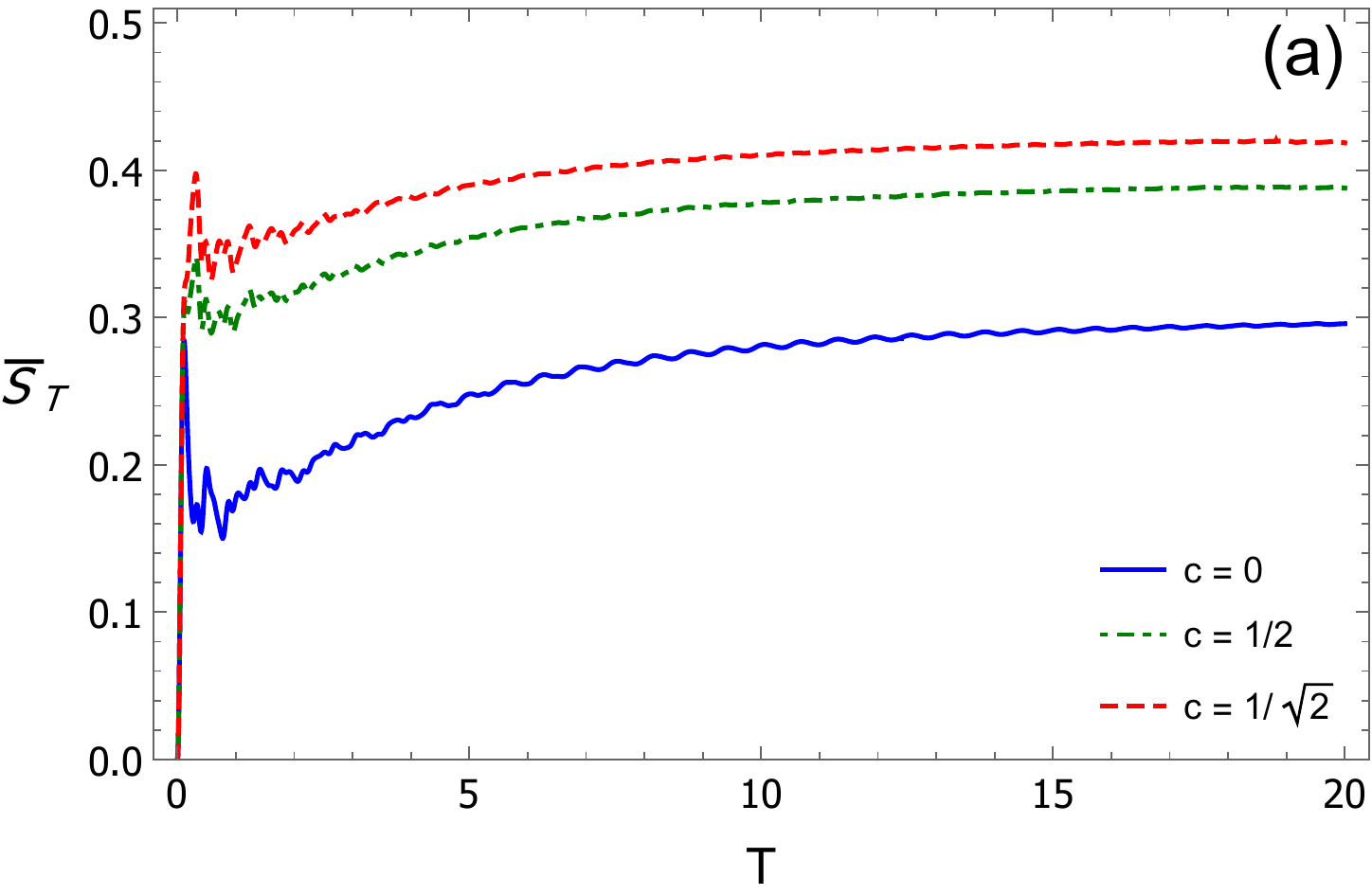}}\qquad\qquad
	\subfigure{\includegraphics[scale=0.40]{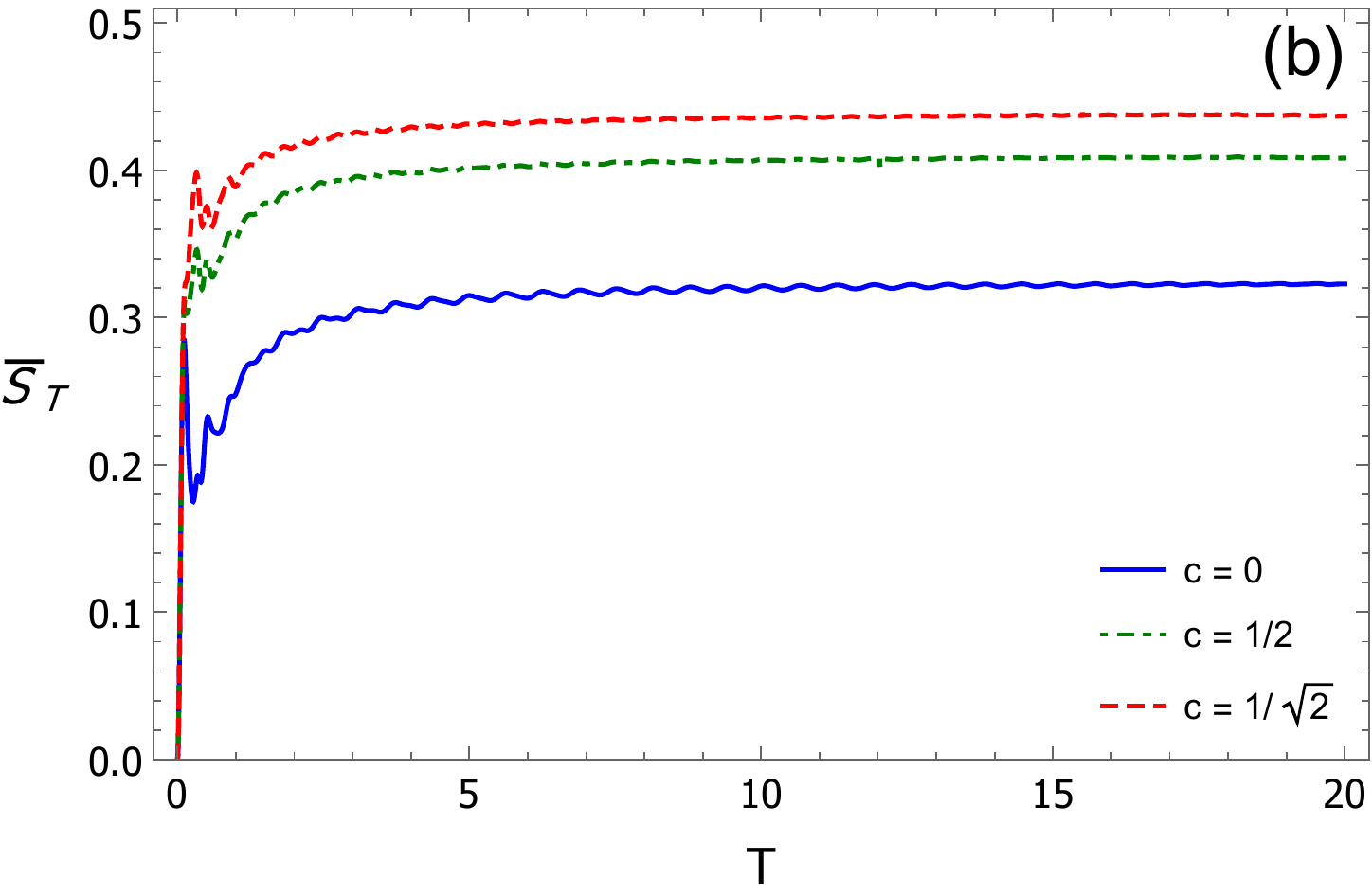}}\qquad
	\caption{Evolution of the cumulative time-average of the linear entropy, $\bar{S}_T$, for: (a) $\theta = 0$ and (b) $\theta = \pi$.
		The continuous (blue), dot-dashed (green), and dashed (red) curves correspond to $c = 0$, $c = 1/2$, and $c = 1/\sqrt{2}$, respectively. }
	\label{linearentropyTA}
\end{figure}

\begin{figure}[htpb]
	\centering
	\includegraphics[scale=0.60]{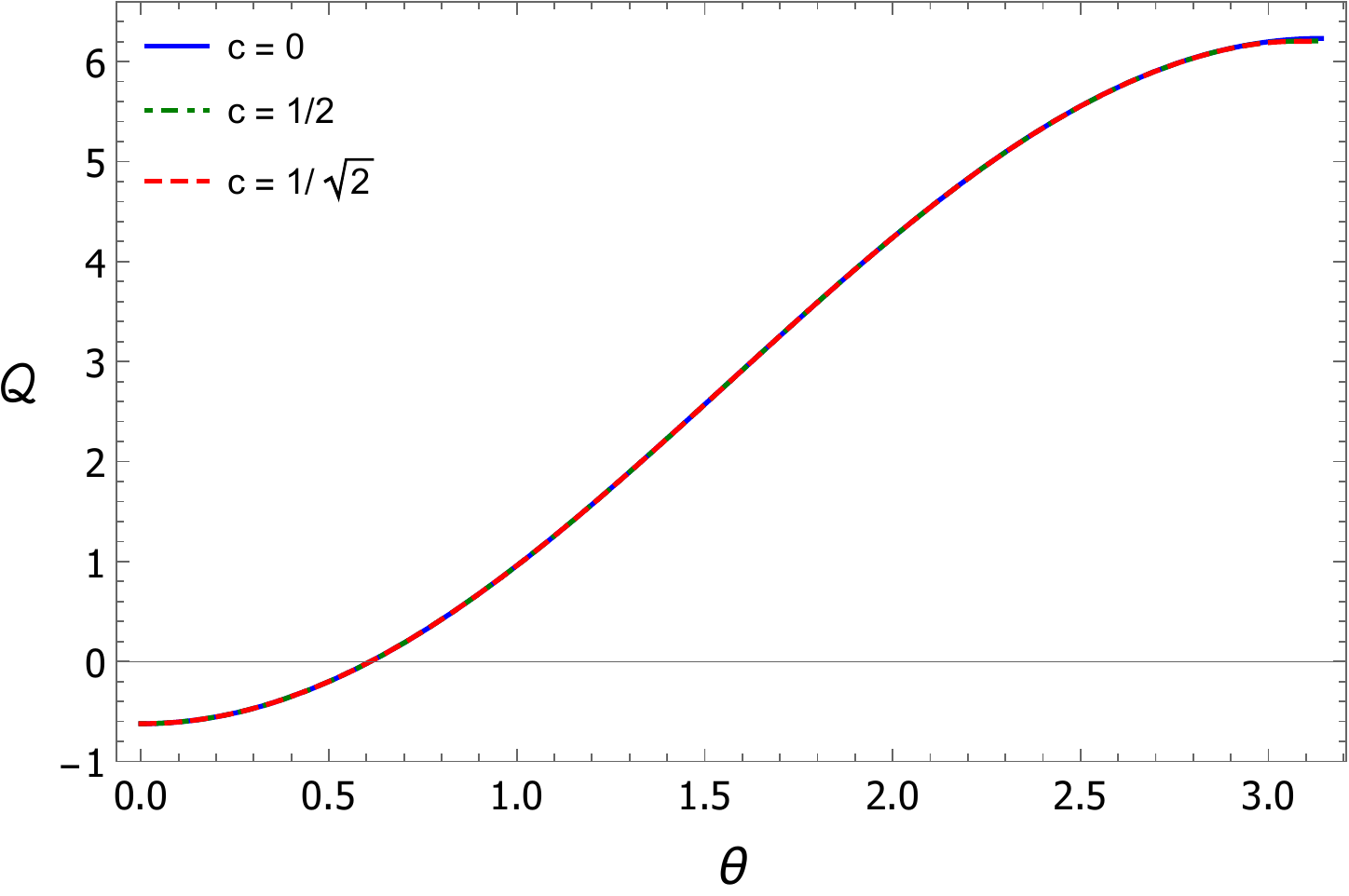}
	\caption{Mandel's $Q$ parameter of squeezed states (and superpositions with $\varphi = 0$) as a function of $\theta$. Depending 
		on the value of $\theta$, the states can be sub-Poissonian, super-Poissonian as well as Poissonian.}
	\label{qmandel}
\end{figure}

In order to have a better understanding of the above results, we calculated the cumulative time-average of the linear entropy, 
$\bar{S}_T$, defined as
\begin{equation}
	\bar{S}_{T}\left(T\right)=\frac{1}{T}\int_{0}^{T}S\left(t\right)dt\,. 
\end{equation}
The results are shown in Fig.(\ref{linearentropyTA}). We notice that the average values of the linear entropy tend to get flat and 
converge to values that are the highest for an environment in an equally-weighed superposition state ($c = 1/\sqrt{2}$). Also,
the values of $\bar{S}_{T}\left(T\right)$ at longer times are slightly higher for $\phi = \pi$ (keeping $c$ fixed), as we see by 
comparing the plots. We should also point out that, despite the oscillatory pattern of the linear entropy, its long-time average seems 
to saturate to constant values. 
The influences of each of these parameters 
is evident in the above-mentioned figures. The state of qubit 1 tends to be more mixed if: i) the environment is initially in an equally-weighed 
superposition state; ii) the states composing the environment are phase squeezed states. As we are going to see below, these results can be understood 
based on the amount of quantum noise present in the initial state of the environment, which depends on the parameters $\theta$ and $c$.
We should stress that although the state in Eq.(\ref{initialfield}) is pure, it might exhibit fluctuations in photon number as well 
as in the quadrature variables.
Firstly we analyze the photon number fluctuations, quantified by Mandel's $Q$ parameter, defined as \cite{mandel79}, 
\begin{equation}
	Q=\frac{\left\langle \left(\Delta\hat{n}\right)^{2}\right\rangle -\left\langle \hat{n}\right\rangle }{\left\langle \hat{n}\right\rangle}.   
\end{equation}
The $Q$ parameter indicates deviations from the Poissonian photon statistics 
characteristic of the coherent states, for which $Q = 0$. If $Q > 0$ ($Q < 0$) the state is called super-Poissonian (sub-Poissonian). Its minimum value 
is $Q = -1$, for states having an exact number of photons, e.g., the $N-$photon Fock state $|N\rangle$. The photon statistics of squeezed states is strongly 
dependent on the phase $\theta$. This is clearly seen in Fig.(\ref{qmandel}), where we have plotted the parameter $Q$ as a function of $\theta$. 
The squeezed states, as well as the type of superpositions we are using here (with $\varphi = 0$), have photon number fluctuations that depend on the phase
$\theta$ but are independent of $c$. Yet, the state of the environment is sub-Poissonian for $\theta \lessapprox 0.6$ and super-Poissonian for
$\theta \gtrapprox 0.6$. The linear entropy of qubit 1 attains larger values (higher mixedness) for an environment exhibiting larger photon number
fluctuations (super-Poissonian, $\theta = \pi$). On the other hand, a less noisy initial state of the environment ($\theta = 0$) results in a 
less mixed state for qubit 1, as verified. We should remark that while the $Q$ parameter of the initial state of the environment does not depend on $c$, 
the average linear entropy $\bar{S}_T$ has a dependence on this parameter. This means that we must look for a different explanation for such behavior. In fact, 
the quantum noise can manifest itself in other variables, e.g., in the quadratures of the field, $\hat{X_{1}}$ and $\hat{X_{2}}$, defined as

\begin{equation}
	\hat{X_{1}}=\frac{1}{2}\left(a+a^{\dagger}\right)\;\;\mbox{and}\;\;\hat{X_{2}}=\frac{1}{2i}\left(a-a^{\dagger}\right).
\end{equation}

\begin{figure}[htpb]
	\centering
	\includegraphics[scale=0.60]{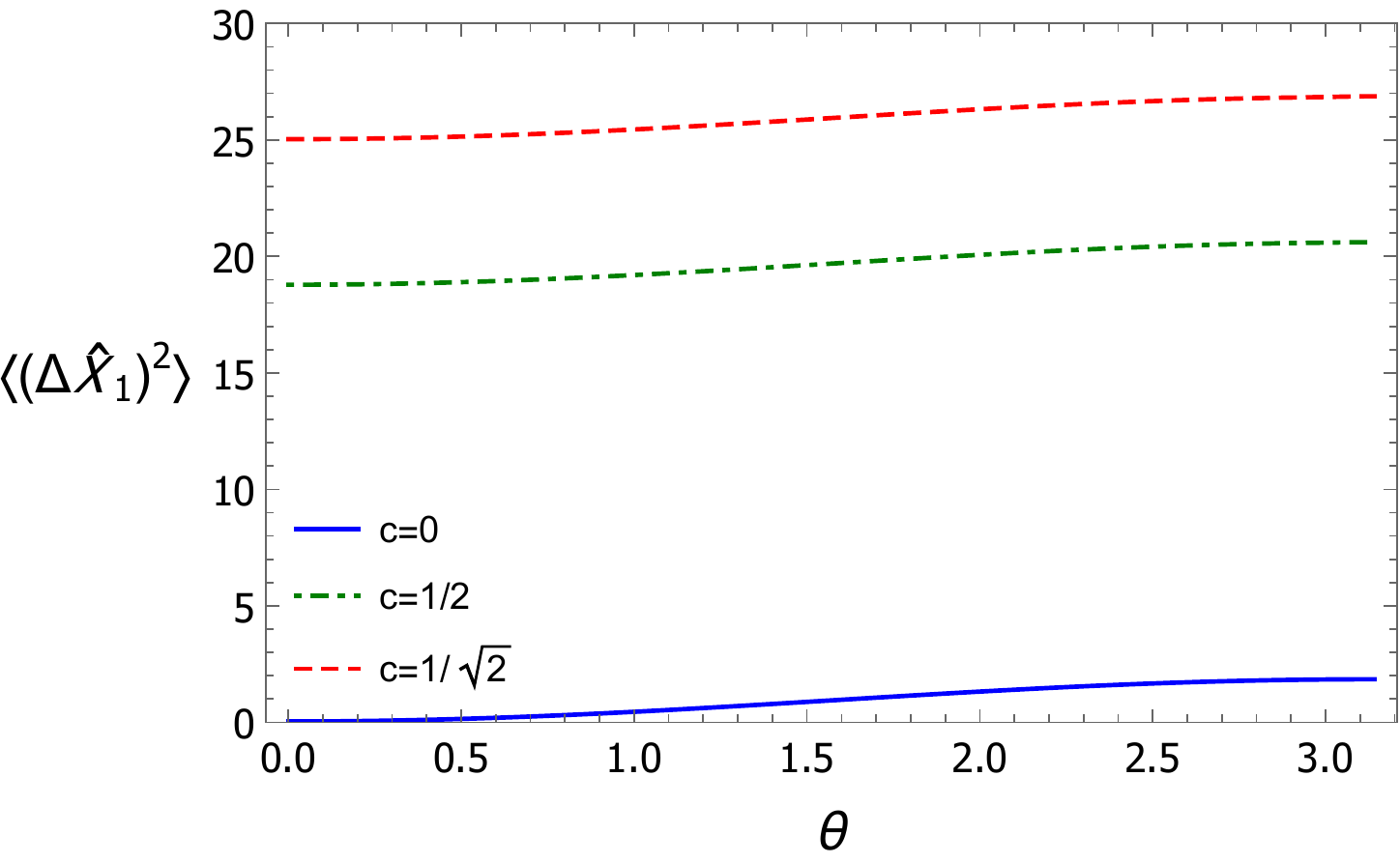}
	\caption{Variance of quadrature $\hat{X_{1}}$, $\langle(\Delta\hat{X_{1}})^{2}\rangle$, of squeezed states 
		(and superpositions with $\varphi = 0$) as a function of $\theta$. The continuous (blue), dot-dashed (green), 
		and dashed (red) curves correspond to $c = 0$, $c = 1/2$, and $c = 1/\sqrt{2}$, respectively. }
	\label{quadratvariance_vs_theta}
\end{figure}

\begin{figure}[htpb]
	\centering
	\includegraphics[scale=0.60]{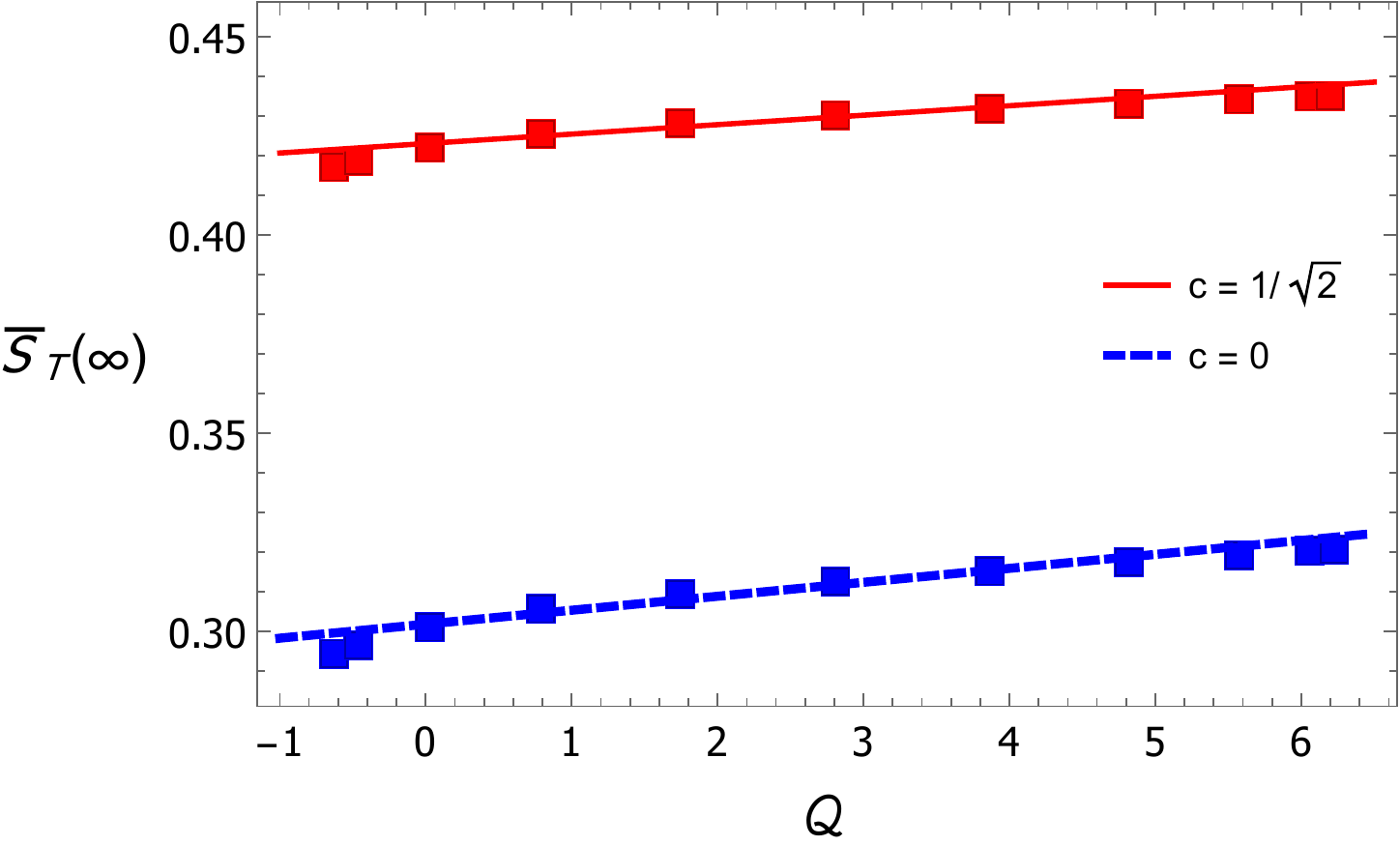}
	\caption{Long-time value of the average linear entropy as a function of Mandel's $Q$ parameter. The continuous (red) curve corresponds to
		$c = 1/\sqrt{2}$ and the dashed (blue) curve to $c = 0$.}
	\label{entropy_vs_Q}
\end{figure}

\begin{figure}[htpb]
	\centering
	\includegraphics[scale=0.60]{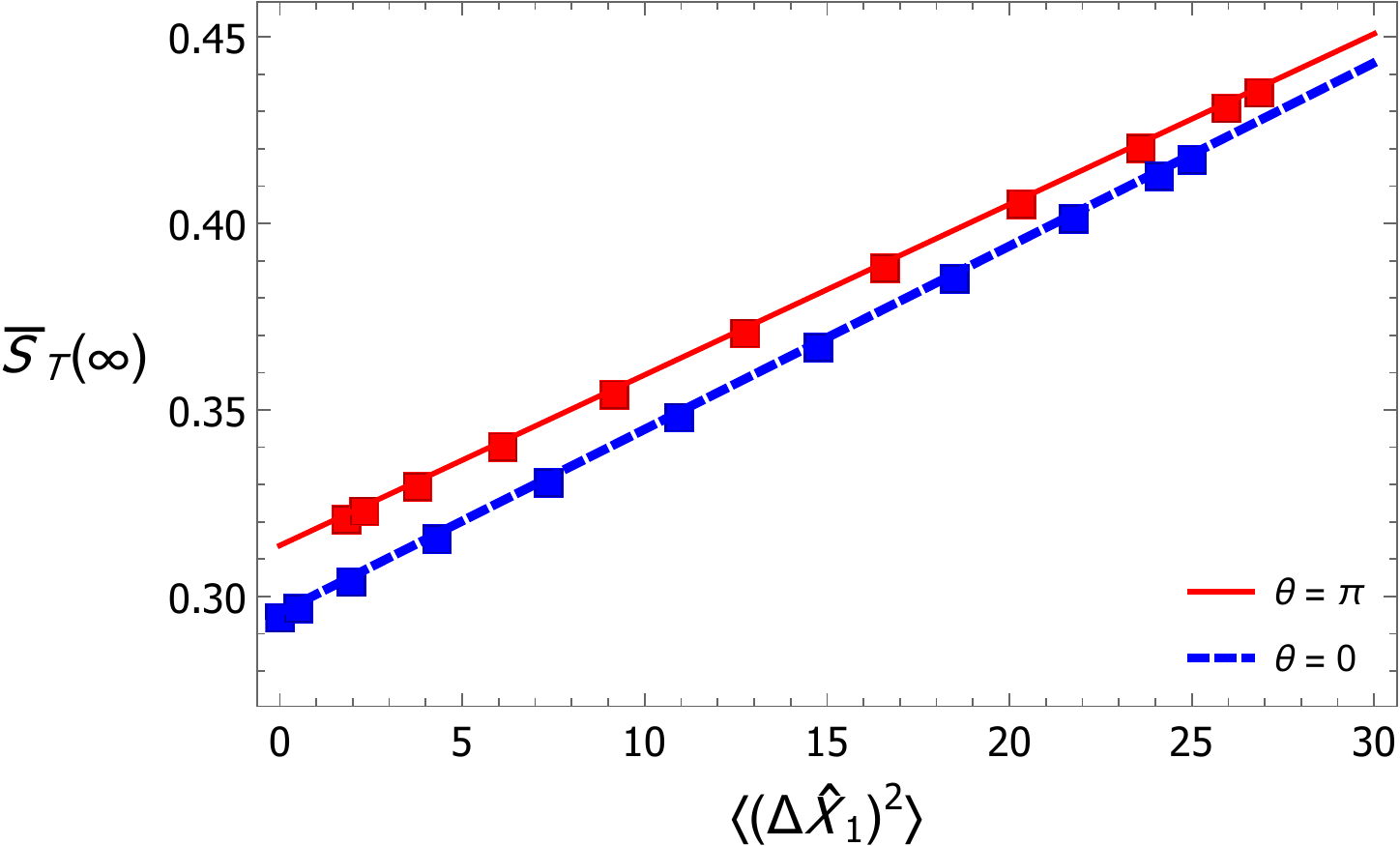}
	\caption{Long-time value of the average linear entropy as a function of the quadrature variance $\langle(\Delta\hat{X_{1}})^{2}\rangle$.
		The continuous (red) curve corresponds to $\theta = \pi$ and the dashed (blue) curve to $\theta = 0$.}
	\label{entropy_vs_quadrature}
\end{figure}

The quadrature operators obey $\left[\hat{X_{1}},\hat{X_{2}}\right]=i/2$, and consequently 
\begin{equation}
	\langle(\Delta\hat{X_{1}})^{2}\rangle \langle\Delta\hat{X_{2}})^{2}\rangle \geq1/16. 
\end{equation}
Squeezing of quantum noise in the $i$-th quadrature is verified if
\begin{equation}
	\langle(\Delta\hat{X_{i}})^{2}\rangle <\frac{1}{4}.
\end{equation}
The superposition of squeezed states constituting the environment can present squeezing or not, depending on the parameter 
$c$. In Fig.(\ref{quadratvariance_vs_theta}) we have plotted the variance of the quadrature $\hat{X_{1}}$, 
$\langle(\Delta\hat{X_{1}})^{2}\rangle$, as a function of $\theta$. We note that the quadrature noise 
assumes the largest values in the equally weighted superposition ($c = 1/\sqrt{2}$). This is in agreement with the fact that 
in this case, the linear entropy of qubit 1 is, on average, also the largest, having fixed the initial mean photon number of the environment. 
Basically, the amount of quantum noise in the initial state of the field, either photon or quadrature noise, directly affects the purity of 
the two-qubit state during the evolution. This is also in agreement with the results considering the environment in a thermal state,
given that in this case Mandel's $Q$ parameter is $Q = \langle\hat{n}\rangle$ and the quadrature variance is 
$\langle(\Delta\hat{X_{1}})^{2}\rangle = (2\langle\hat{n}\rangle+1)/4$, i.e., the larger $\langle\hat{n}\rangle$, the 
noisier will be the thermal state. In order to establish a more quantitative connection between the noise in the environment
and the dynamics of the system, we evaluated numerically the long-time value of the average linear entropy and plotted it against 
Mandel's $Q$ parameter, in Fig.(\ref{entropy_vs_Q}) as well as against the variance of the quadrature $\hat{X_{1}}$, in Fig.(\ref{entropy_vs_quadrature}).
In Fig.(\ref{entropy_vs_Q}) we used $11$ values of $\theta$ from $\theta = 0$ to $\theta = \pi$, whereas in Fig.(\ref{entropy_vs_quadrature}) 
we used also $11$ values of $c$, from $c = 0$ to $c = 1/\sqrt{2}$ (squares in the plots in both figures). Interestingly, we verified a behaviour that is
very close to a linear relation between the long-term average linear entropy and the fluctuations in the initial state of the environment for both photon 
and variance noise cases, despite the fact that they are distinct types of noise.
\subsection{Evolution of the qubit 1-qubit 2 quantum entanglement}
\label{section022}

\begin{figure}[htpb]
	\centering
	\subfigure{\includegraphics[scale=0.40]{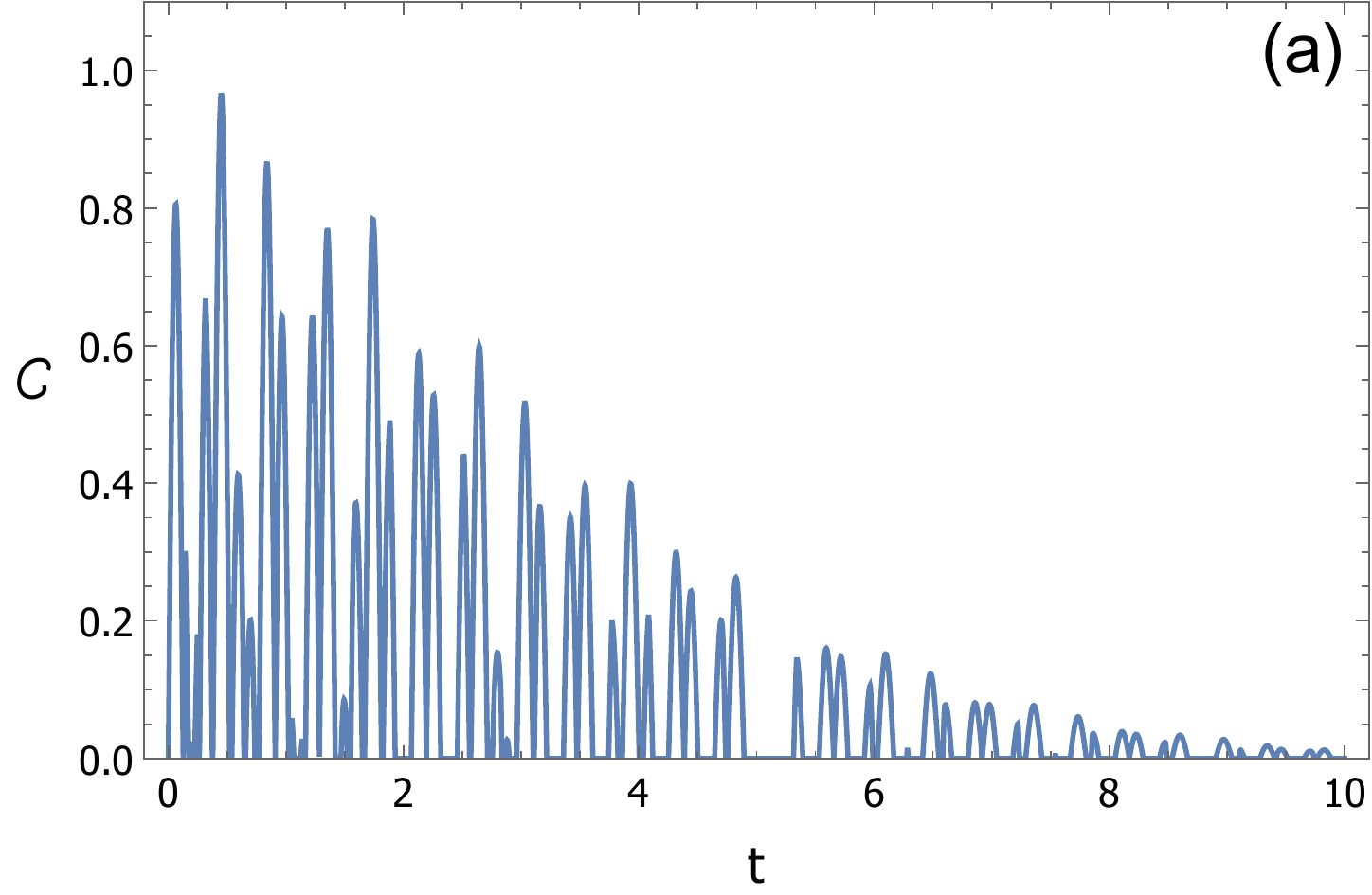}}\qquad\qquad
	\subfigure{\includegraphics[scale=0.40]{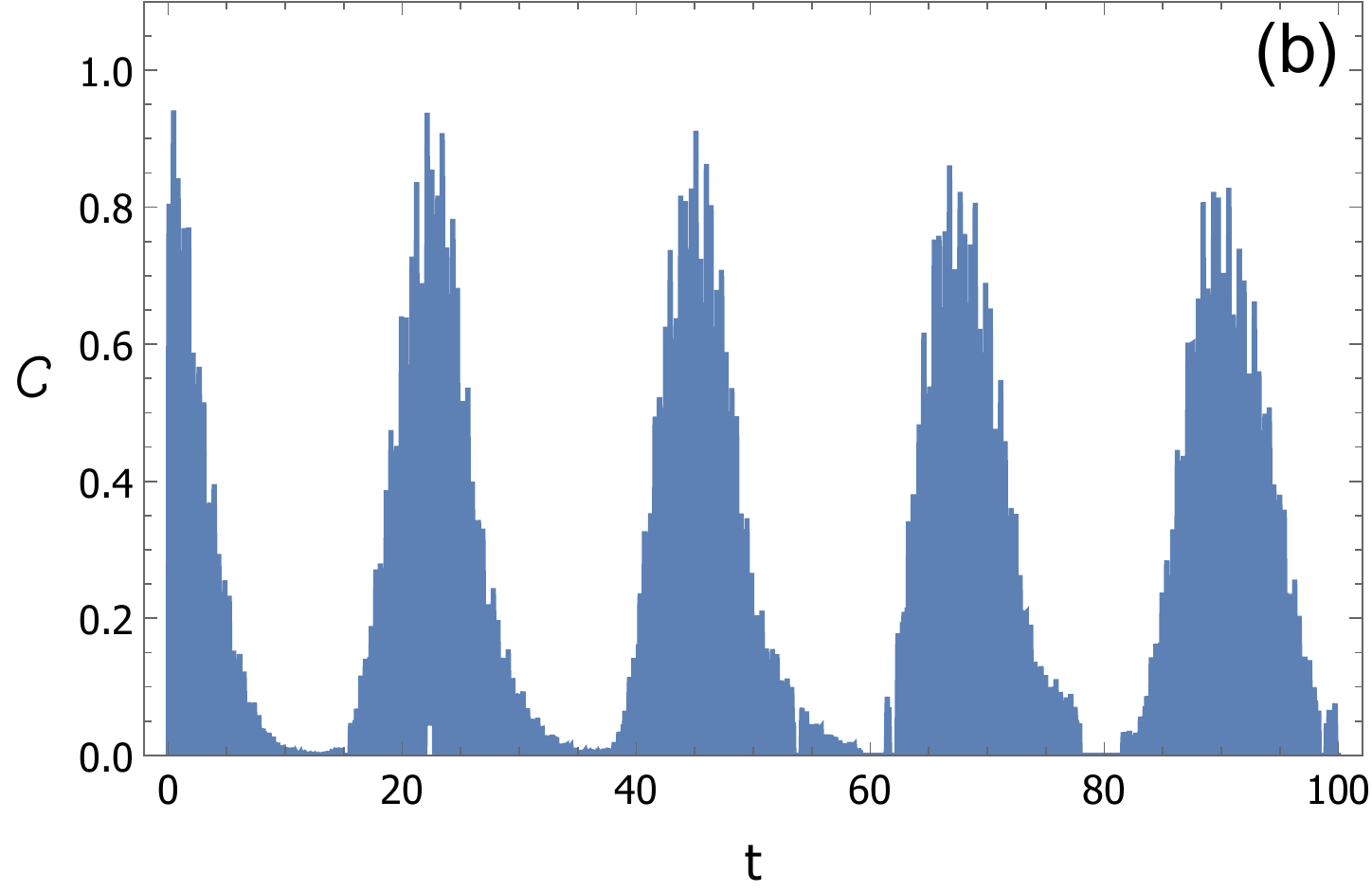}}\qquad
	\caption{Concurrence of the two qubits as a function of time, for $\theta = 0$ (amplitude squeezing) and $c = 1/\sqrt{2}$ 
		(equally weighted superposition) on a short time-scale (a), and a long time-scale (b). 
		Here, $r = 1.0$, $\alpha = 5.0$ and $\varphi = \pi$.}
	\label{concurrence_theta_zero}
\end{figure}

\begin{figure}[htpb]
	\centering
	\subfigure{\includegraphics[scale=0.40]{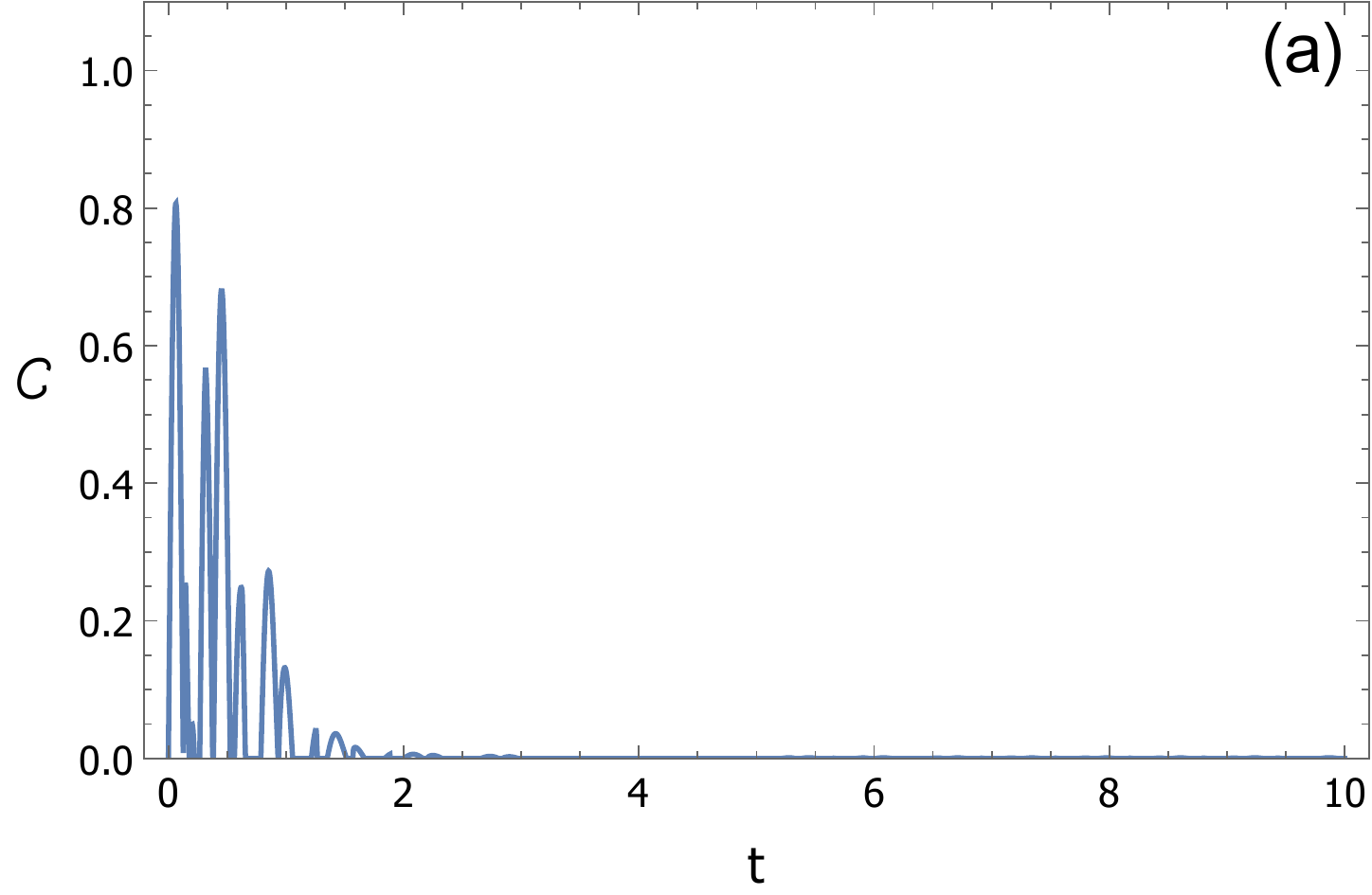}}\qquad\qquad
	\subfigure{\includegraphics[scale=0.40]{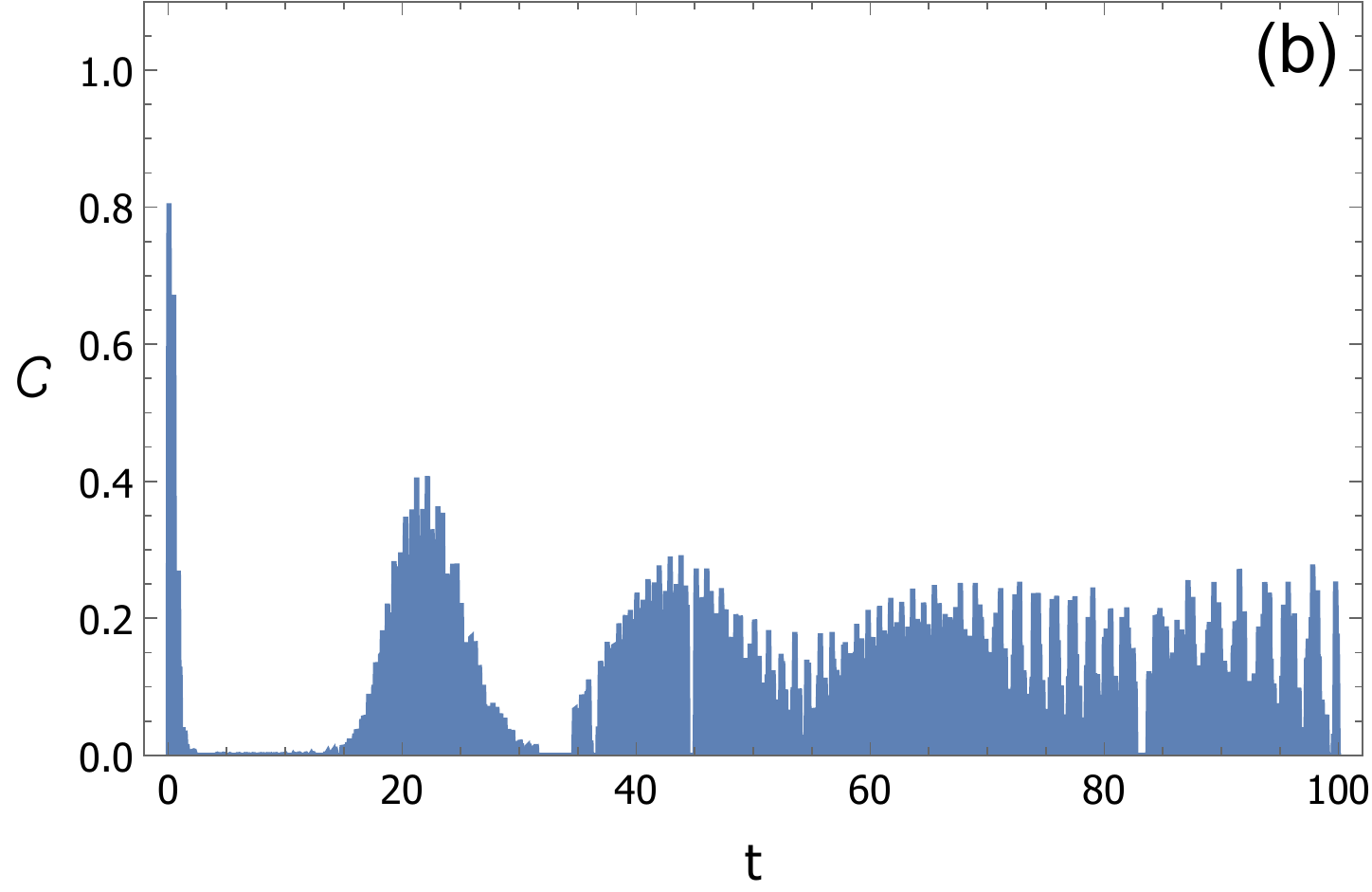}}\qquad
	\caption{Concurrence of the two qubits as a function of time, for $\theta = \pi$ (phase squeezing) and $c = 1/\sqrt{2}$ 
		(equally weighted superposition) on a short time-scale (a), and a long time-scale (b). Here, 
		$r = 1.0$, $\alpha = 5.0$ and $\varphi = \pi$.}
	\label{concurrence_theta_pi}
\end{figure}

Quantum entanglement is without any doubt one of the most striking quantum effects that can also be a resource to perform
certain tasks \cite{horodecki09}. As mentioned, we have already shown that a thermal, phase-insensitive environment can induce the
Entanglement Sudden Death/Birth effects in such a two-qubit system \cite{decordi20}. Here, we report similar effects in a 
scenario involving a phase-sensitive environment. The qubit-qubit interaction naturally leads to bipartite entanglement in 
this system, that can be quantified in a straightforward way, e.g., via the concurrence ${\cal C}$ \cite{wootters97}. 
In our model involving two qubits, we may compute ${\cal C}(t)$, the concurrence as a function of time, as follows 

\begin{equation}
	\mathcal{C\mathrm{\left(t\right)}=\mathrm{max\,\left[0,\Lambda\left(t\right)\right]\,,}}
\end{equation}
with 

\begin{equation}
	\Lambda\left(t\right)\equiv\sqrt{\xi_{1}\left(t\right)}-\sqrt{\xi_{2}\left(t\right)}-\sqrt{\xi_{3}\left(t\right)}-\sqrt{\xi_{4}\left(t\right)}\,.
\end{equation}
The quantities $\xi_{i}$ are the eigenvalues of the matrix 
\begin{equation}
	M\left(t\right)=\rho_{q1,q2}\left(t\right)\left(\sigma_{y}^{(1)}
	\otimes\sigma_{y}^{(2)}\right)\rho^{*}_{q1,q2}\left(t\right)\left(\sigma_{y}^{(1)}\otimes\sigma_{y}^{(2)}\right),
\end{equation}
which should be placed in decreasing order, and $\sigma_{y}^{(i)}$ is the Pauli matrix of the $i$-th qubit. In what follows we present 
some plots of the time evolution of the bipartite quantum entanglement of the 2 qubits for the initial state of the environment in a 
superposition state with $c = 1/\sqrt{2}$. In Fig.(\ref{concurrence_theta_zero}) we have 
plots of the concurrence of the two-qubit system as a function of time for $\theta = 0$. We note in Fig.(\ref{concurrence_theta_zero})a that for short times,
there are decaying oscillations of the concurrence and also brief intervals in which the concurrence is zero (Entanglement Sudden Death). Yet, for longer 
times there are almost periodic revivals of quantum entanglement, as seen in Fig.(\ref{concurrence_theta_zero})b. On the other hand, if $\theta = \pi$, the 
increased fluctuations in the environment result in a larger degradation of quantum entanglement. We notice longer time intervals of Entanglement Sudden Death 
[see Fig.(\ref{concurrence_theta_pi})], while as time passes, the concurrence has, on average, lower values than in the previous case ($\theta = 0$).
\subsection{Evolution of the two-qubit quantum coherence}
\label{section023}

Quantum coherence plays a central role in quantum theory; not only can be quantified \cite{plenio14}, but it also represents a 
resource \cite{plenio17} useful for performing quantum information tasks. An intuitive and convenient way of quantifying quantum coherence is
via the $l_1$-norm of coherence \cite{plenio14}, defined as
\begin{equation}
	C_{l_{1}}=\sum_{i,j\,,i\neq j}\left|\rho_{ij}\right|,
\end{equation}
being $\rho_{ij} \equiv \left\langle i\right|\rho_{q}\left|j\right\rangle$ the matrix elements relative to the considered system.
We calculated the $l_1$-norm of coherence as a function of time in order to quantify the quantum coherence of the 2 qubits, a joint 
property of the system. The result is
\begin{eqnarray}
	C_{l_{1}}(t) & = & 2\mathcal{N}^{2}\left[\left|\sum_{n=0}^{\infty}\gamma_{n+1}\gamma_{n}^{*}A_{12}^{\left(n+1\right)}A_{22}^{\left(n\right)*}\right|
	+\left|\sum_{n=0}^{\infty}\gamma_{n+1}\gamma_{n}^{*}A_{12}^{\left(n+1\right)}A_{23}^{\left(n\right)*}\right|\right. \\ 
	&+&\left.\left|\sum_{n=0}^{\infty}\gamma_{n+2}\gamma_{n}^{*}A_{12}^{\left(n+2\right)}A_{24}^{\left(n\right)*}\right|
	+\left|\sum_{n=0}^{\infty}\gamma_{n}\gamma_{n}^{*}A_{22}^{\left(n\right)}A_{23}^{\left(n\right)*}\right|\right. \\ 
	&+& \left.\left|\sum_{n=0}^{\infty}\gamma_{n+1}\gamma_{n}^{*}A_{22}^{\left(n+1\right)}A_{24}^{\left(n\right)*}\right|
	+\left|\sum_{n=0}^{\infty}\gamma_{n+1}\gamma_{n}^{*}A_{23}^{\left(n+1\right)}A_{24}^{\left(n\right)*}\right|\right],
\end{eqnarray}
where the coefficients $\gamma_n$ and $A_{ij}$ can be found in the Appendix.

\begin{figure}[htpb]
	\centering
	\subfigure{\includegraphics[scale=0.40]{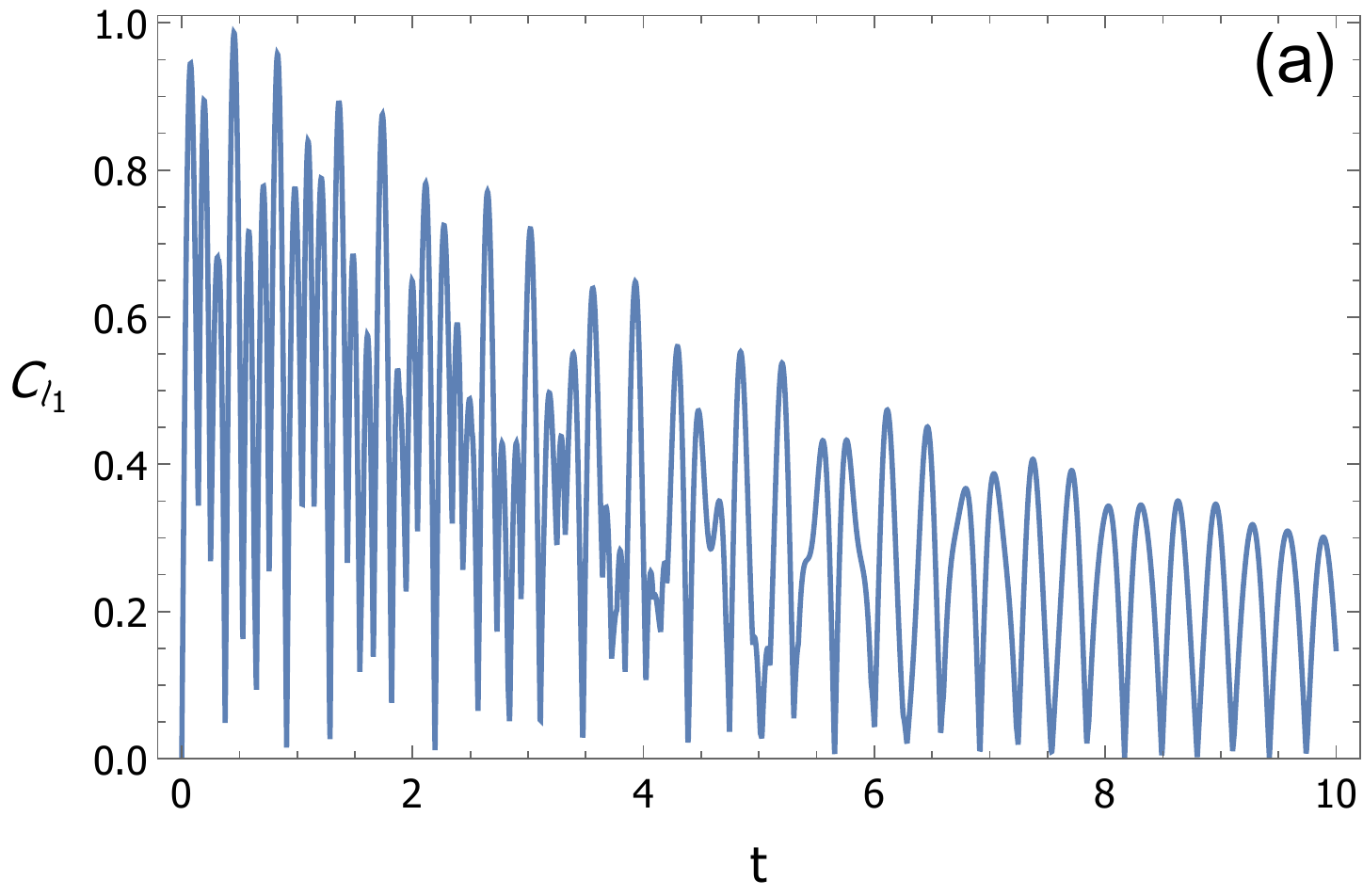}}\qquad\qquad
	\subfigure{\includegraphics[scale=0.40]{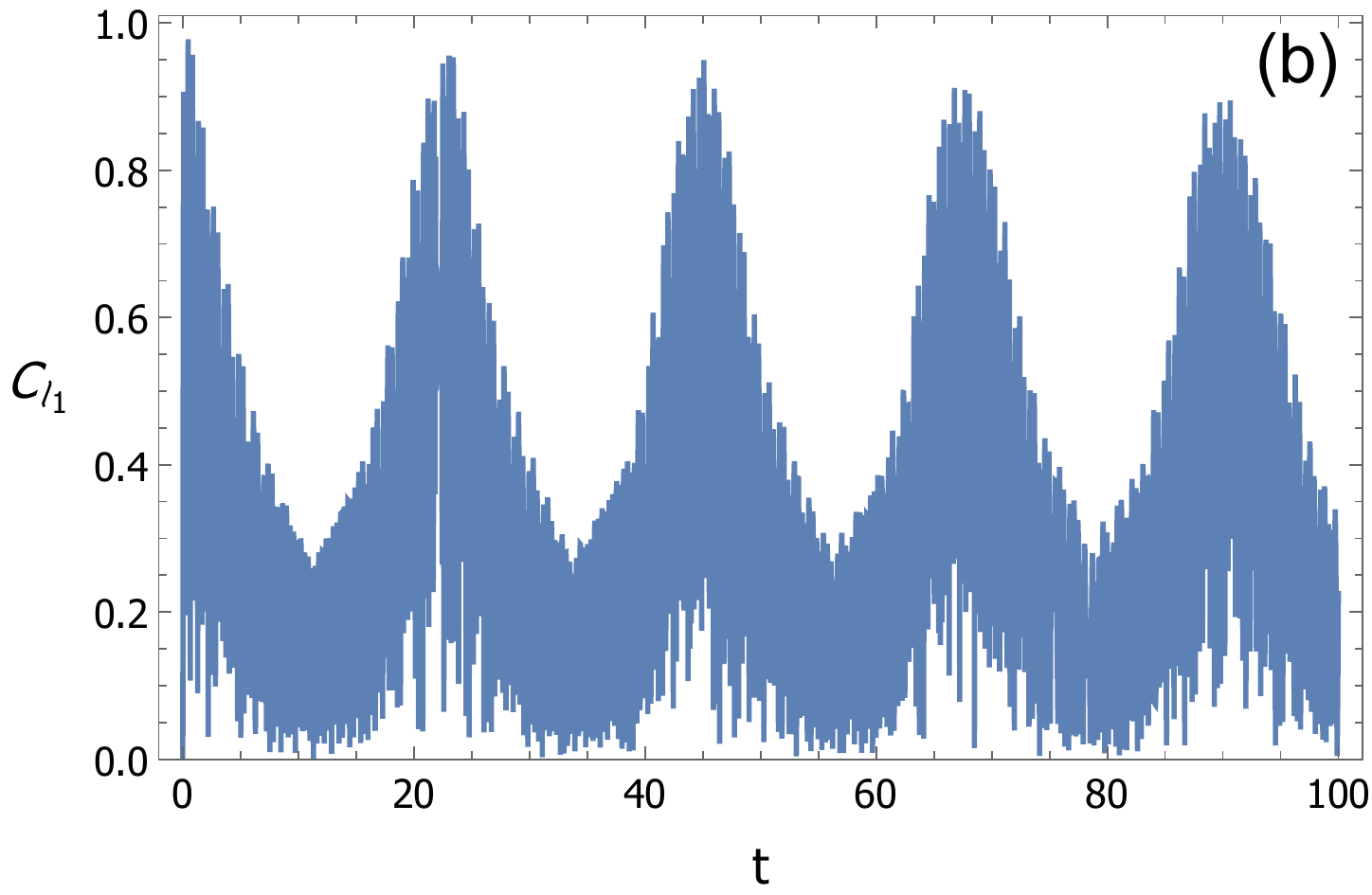}}\qquad
	\caption{Quantum coherence of the two qubits as a function of time, for $\theta = 0$ (amplitude squeezing) and $c = 1/\sqrt{2}$ 
		(equally weighted superposition) on a short time-scale (a), and a long time-scale (b). Here, 
		$r = 1.0$, $\alpha = 5.0$ and $\varphi = \pi$.}
	\label{coherence_theta_0}
\end{figure}

\begin{figure}[htpb]
	\centering
	\subfigure{\includegraphics[scale=0.40]{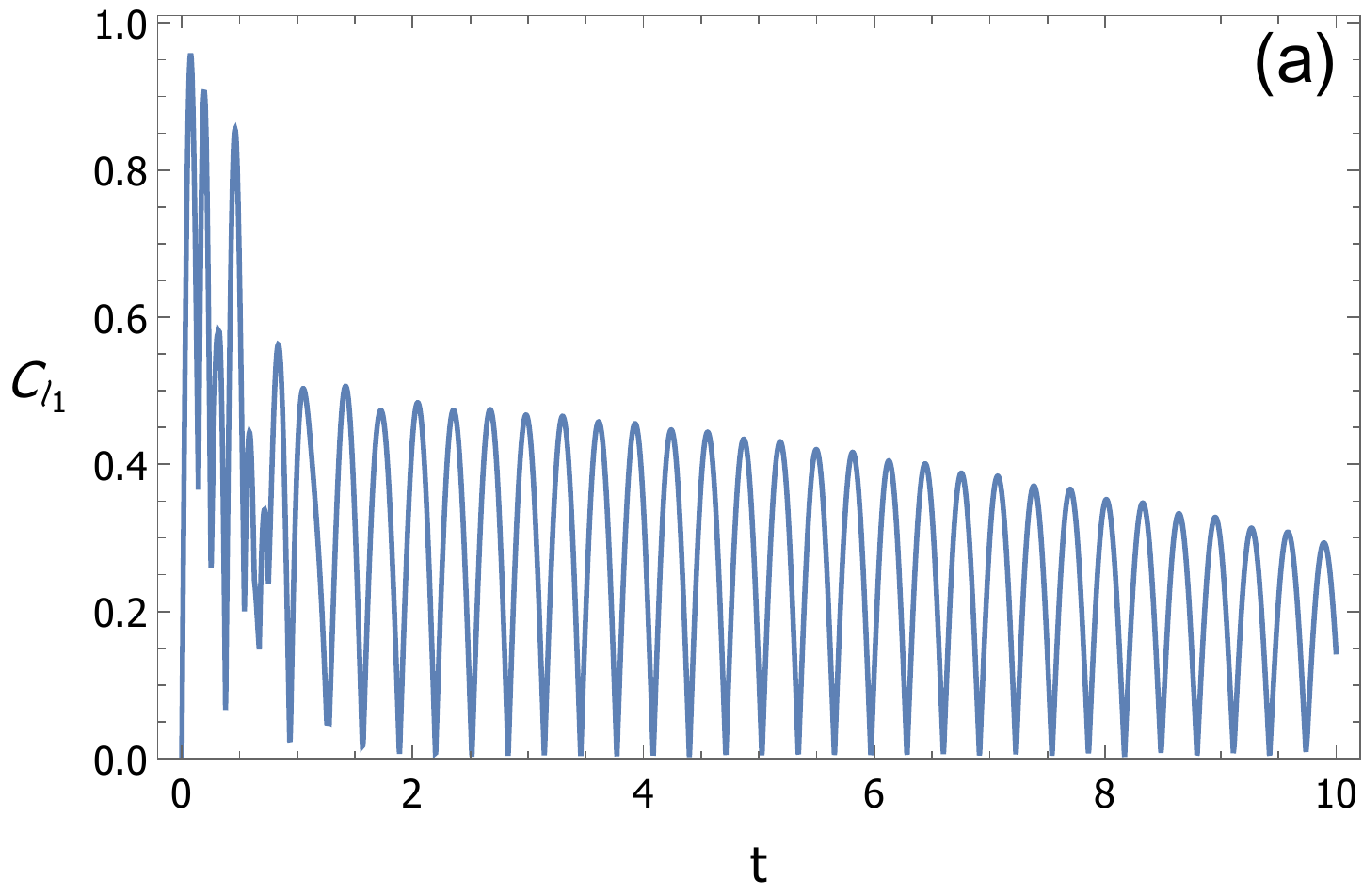}}\qquad\qquad
	\subfigure{\includegraphics[scale=0.40]{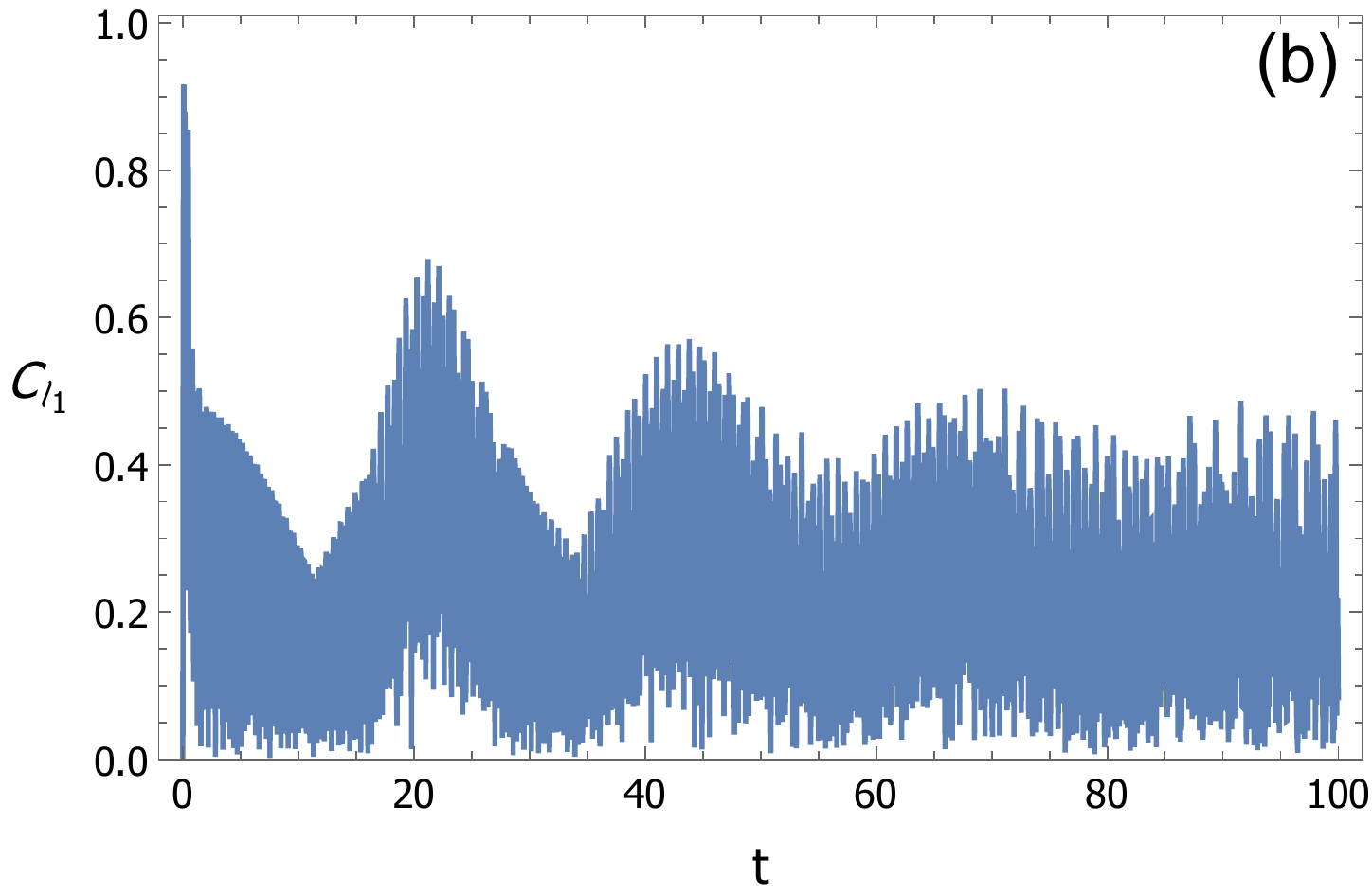}}\qquad
	\caption{Quantum coherence of the two qubits as a function of time, for $\theta = \pi$ (phase squeezing) and $c = 1/\sqrt{2}$ 
		(equally weighted superposition) on a short time-scale (a), and a long time-scale (b). Here, 
		$r = 1.0$, $\alpha = 5.0$ and $\varphi = \pi$.}
	\label{coherence_theta_pi}
\end{figure}

Because of the specific initial conditions we have chosen here, the two-qubit system has zero initial quantum coherence, i.e., 
$C_{l_{1}}(t=0)$ = 0.0. Due to the qubit-qubit interaction, coherence is basically an oscillatory function of time if the system is isolated,
ranging from $0.0$ to $1.0$. Yet, such regularity is disrupted due to the coupling with the small environment. 
We show now the results for $\theta = 0$, in Fig.\ref{coherence_theta_0},  and for $\theta = \pi$, in Fig.\ref{coherence_theta_pi},
for both short and long time-scales and $c =  1/\sqrt{2}$. We note that the system acquires a certain quantum coherence at the beginning of evolution, 
which is then degraded due to the interaction with the environment. Furthermore, the loss of coherence is more pronounced in the 
case of a noisier initial environment, that is, for $\theta = \pi$, as we see comparing the results. This is consistent with the previously obtained
results, regarding the time-evolution of the linear entropy and entanglement. 
\newpage
\section{Conclusions}

We have studied the dynamics of a two-qubit system under the influence of a phase-sensitive, small environment constituted by a
single-mode field. In our model, the two qubits (qubit 1 and qubit 2) are coupled, but only one of them (qubit 2) is interacting with 
the environment.  We obtained an analytical solution to the problem and assumed that the field was initially in a pure state, a quantum superposition of 
squeezed coherent states. After tracing over the environment variables, we have calculated numerically quantities such as the linear entropy of qubit 1 
as well as the concurrence (quantum entanglement) relative to the two-qubit system. Those quantities are normally 
periodic functions of time in completely isolated systems, but the coupling to a third system (environment) may cause disturbances to such regular 
evolution, even if the environment consists of a single-mode field. The studied quantities still oscillate in time in the presence of such an 
environment, but now start having varying amplitudes. The initial state of the field, despite being pure, may exhibit fluctuations in both photon 
number and quadrature variables, which has a significant impact on the evolution of the two-qubit system. Also, a variety of quantum states can be
approximated by the state in Eq.(\ref{initialfield}), which allows the mimicking of different types of environments depending on the phase $\theta$ 
and the weight in the superposition, $c$. The small environment we consider here is by no means in a thermal equilibrium state, 
but we still were able to identify a very weak form of equilibration in the sense that the cumulative time average of the linear entropy, $\bar{S}_T$, 
seemingly tends to a constant value for relatively long times. We have also found an interesting relationship regarding the linear entropy; a virtually 
linear relation between the long time value of $\bar{S}_T$ and the fluctuations present in the initial state of the environment. 
This work can be useful for investigations of scenarios and control of systems of interest that are subjected to environments
having a small number of degrees of freedom.

\appendix

\section{Analytical solution of the model: two coupled qubits interacting with a superposition of squeezed states}

Consider the model of two coupled qubits interacting with a single mode field described by the Hamiltonian in 
Eqs.(\ref{hamiltot}), (\ref{hamilqubits}) and (\ref{hamilqubitfield}), with initial joint state
$$
|\Psi(0)\rangle = \left|e_{1}\right\rangle \otimes \left|g_{2}\right\rangle \otimes |\psi_{f}(0)\rangle,
$$
being $|\psi_{f}(0)\rangle$ the initial field state in Eq.(\ref{initialfield}). The state $|\Psi(0)\rangle$ can 
be recast in the form
$$
\left|\Psi\left(0\right)\right\rangle =\mathcal{N}\sum_{n=0}^{\infty}\gamma_{n}\left(\alpha,\,r,\,\theta\right)\left|e_{1},\,g_{2},\,n\right\rangle,
$$
where 
$$
\gamma_{n}\left(\alpha,\,r,\,\theta\right)\equiv c\,C_{n}\left(\alpha,\,r,\,\theta\right)+e^{i\,\varphi}\sqrt{1-c^{2}}\,C_{n}\left(-\alpha,\,r,\,\theta\right).
$$
Here 
$$
C_{n}\left(\alpha,\,r,\,\theta\right)=\frac{\left(\frac{\nu}{2\mu}\right)^{\frac{n}{2}}}{\sqrt{n!\,\mu}}\exp\left[\frac{-\left|\alpha\right|^{2}}{2}-\frac{1}{2}\alpha^{*2}e^{i\,\theta}\tanh r\right]H_{n}\left(\frac{\alpha\mu+\alpha^{*}\nu}{\sqrt{e^{i\,\theta}\sinh 2r}}\right)\nonumber
$$
is the coefficient of the squeezed coherent state in the Fock basis, that is, 
$\left|\alpha,\,\xi\right\rangle =\sum_{n=0}^{\infty}C_{n}\left(\alpha,\,r,\,\theta\right)\left|n\right\rangle$.

The solution of the model, that is, the system's (2 qubits $+$ field) time-evolved state vector in the interaction representation, 
$|\Psi(t)\rangle_I = e^{-i\,H_{1}t}|\Psi(0)\rangle$, is given by
\begin{eqnarray*}
	|\Psi (t)\rangle_{I} & = &
	\mathcal{N}\sum_{n=0}^{\infty}\gamma_{n} (\alpha,\,r,\,\theta) [A_{12}^{ (n)} (t) |e_{1},\,e_{2},\,n-1\rangle +A_{22}^{(n)} (t) 
	|e_{1},\,g_{2},\,n\rangle + \\
	& + &A_{23}^{ (n)} (t) |g_{1},\,e_{2},\,n\rangle +A_{24}^{(n)} (t) |g_{1},\,g_{2},\,n+1\rangle],
\end{eqnarray*}
where the coefficients $A_{ij}^{ (n)} (t)$ are
\begin{eqnarray*}
	A_{12}^{\left(n\right)}\left(t\right) & = & \frac{i\,a_{n}}{r_{n}}\left[\frac{\left(b_{n}^{2}-\omega_{+,n}^{2}\right)}{\omega_{+,n}}\sin(\omega_{+,n}t)-\frac{\left(b_{n}^{2}
		-\omega_{-,n}^{2}\right)}{\omega_{-,n}}\sin(\omega_{-,n}t)\right] \\
	A_{22}^{\left(n\right)}\left(t\right) & = & \frac{1}{r_{n}}\left[\left(\omega_{+,n}^{2}-b_{n}^{2}\right)\cos(\omega_{+,n}t)-\left(\omega_{-,n}^{2}-b_{n}^{2}\right)
	\cos(\omega_{-,n}t)\right] \\
	A_{23}^{\left(n\right)}\left(t\right) & = & -\frac{i\,\lambda}{r_{n}}\left[\omega_{+,n}\sin(\omega_{+,n}t)-\omega_{-,n}\sin(\omega_{-,n}t)\right] \\
	A_{24}^{\left(n\right)}\left(t\right) & = & \frac{\lambda\,b_{n}}{r_{n}}\left[\cos(\omega_{+,n}t)-\cos(\omega_{-,n}t)\right],
\end{eqnarray*}
with
$$
a_{n}=g\sqrt{n}\,,\qquad b_{n}=g\sqrt{n+1}\,,\label{eq:an+bn}
$$

$$
r_{n}=\sqrt{\left(g^{2}+\lambda^{2}\right)^{2}+4ng^{2}\lambda^{2}}\,,\label{eq:rn}
$$
and
$$
\omega_{\pm,n}=\frac{1}{\sqrt{2}}\sqrt{\left(2n+1\right)g^{2}+\lambda^{2}\pm\sqrt{\left(g^{2}+\lambda^{2}\right)^{2}+4ng^{2}\lambda^{2}}}\,.\label{eq:omg}
$$
From this result, we can obtain the reduced density operator for the two-qubit system, $\rho_{q}\left(t\right)$, by tracing over the field variables,
$\rho_{q}\left(t\right) = \mbox{Tr}_{f}\Big[|\Psi (t)\rangle_{I}{}_{I}\langle \Psi (t) |\Big]$:

\begin{eqnarray*}
	\rho_{q}\left(t\right) =\mathcal{N}^{2}\left[\sum_{n=0}^{\infty}\left|\gamma_{n+1}\right|^{2}\left|A_{12}^{\left(n+1\right)}\right|^{2}\left|e_{1},\,e_{2}\right\rangle \left\langle e_{1},\,e_{2}\right|+\sum_{n=0}^{\infty}\gamma_{n+1}\gamma_{n}^{*}A_{12}^{\left(n+1\right)}A_{22}^{\left(n\right)*}\left|e_{1},\,e_{2}\right\rangle \left\langle e_{1},\,g_{2}\right|\right.\\
	+\sum_{n=0}^{\infty}\gamma_{n+1}\gamma_{n}^{*}A_{12}^{\left(n+1\right)}A_{23}^{\left(n\right)*}\left|e_{1},\,e_{2}\right\rangle \left\langle g_{1},\,e_{2}\right|+\sum_{n=0}^{\infty}\gamma_{n+2}\gamma_{n}^{*}A_{12}^{\left(n+2\right)}A_{24}^{\left(n\right)*}\left|e_{1},\,e_{2}\right\rangle \left\langle g_{1},\,g_{2}\right|\\
	+\sum_{n=0}^{\infty}\gamma_{n}\gamma_{n+1}^{*}A_{22}^{\left(n\right)}A_{12}^{\left(n+1\right)*}\left|e_{1},\,g_{2}\right\rangle \left\langle e_{1},\,e_{2}\right|+\sum_{n=0}^{\infty}\left|\gamma_{n}\right|^{2}\left|A_{22}^{\left(n\right)}\right|^{2}\left|e_{1},\,g_{2}\right\rangle \left\langle e_{1},\,g_{2}\right|\\
	+\sum_{n=0}^{\infty}\gamma_{n}\gamma_{n}^{*}A_{22}^{\left(n\right)}A_{23}^{\left(n\right)*}\left|e_{1},\,g_{2}\right\rangle \left\langle g_{1},\,e_{2}\right|+\sum_{n=0}^{\infty}\gamma_{n+1}\gamma_{n}^{*}A_{22}^{\left(n+1\right)}A_{24}^{\left(n\right)*}\left|e_{1},\,g_{2}\right\rangle \left\langle g_{1},\,g_{2}\right|\\
	+\sum_{n=0}^{\infty}\gamma_{n}\gamma_{n+1}^{*}A_{23}^{\left(n\right)}A_{12}^{\left(n+1\right)*}\left|g_{1},\,e_{2}\right\rangle \left\langle e_{1},\,e_{2}\right|+\sum_{n=0}^{\infty}\gamma_{n}\gamma_{n}^{*}A_{23}^{\left(n\right)}A_{22}^{\left(n\right)*}\left|g_{1},\,e_{2}\right\rangle \left\langle e_{1},\,g_{2}\right|\\
	+\sum_{n=0}^{\infty}\left|\gamma_{n}\right|^{2}\left|A_{23}^{\left(n\right)}\right|^{2}\left|g_{1},\,e_{2}\right\rangle \left\langle g_{1},\,e_{2}\right|+\sum_{n=0}^{\infty}\gamma_{n+1}\gamma_{n}^{*}A_{23}^{\left(n+1\right)}A_{24}^{\left(n\right)*}\left|g_{1},\,e_{2}\right\rangle \left\langle g_{1},\,g_{2}\right|\\
	+\sum_{n=0}^{\infty}\gamma_{n}\gamma_{n+2}^{*}A_{24}^{\left(n\right)}A_{12}^{\left(n+2\right)*}\left|g_{1},\,g_{2}\right\rangle \left\langle e_{1},\,e_{2}\right|+\sum_{n=0}^{\infty}\gamma_{n}\gamma_{n+1}^{*}A_{24}^{\left(n\right)}A_{22}^{\left(n+1\right)*}\left|g_{1},\,g_{2}\right\rangle \left\langle e_{1},\,g_{2}\right|\\
	\left.+\sum_{n=0}^{\infty}\gamma_{n}\gamma_{n+1}^{*}A_{24}^{\left(n\right)}A_{23}^{\left(n+1\right)*}\left|g_{1},\,g_{2}\right\rangle \left\langle g_{1},\,e_{2}\right|+\sum_{n=0}^{\infty}\left|\gamma_{n}\right|^{2}\left|A_{24}^{\left(n\right)}\right|^{2}\left|g_{1},\,g_{2}\right\rangle \left\langle g_{1},\,g_{2}\right|\right].
\end{eqnarray*}

Finally, the reduced density operator relative to qubit 1 is now obtained by tracing over the qubit 2 variables, 
$\rho_{q1}\left(t\right)=\mbox{Tr}_{q2}\left[\rho_{q}\left(t\right)\right]$:
$$
\rho_{q1}\left(t\right)=\rho_{ee}\left|e_{1}\right\rangle \left\langle e_{1}\right|+\rho_{gg}\left|g_{1}\right\rangle \left\langle g_{1}\right|+\rho_{eg}\left|e_{1}\right\rangle \left\langle g_{1}\right|+\rho_{ge}\left|g_{1}\right\rangle \left\langle e_{1}\right|,
$$
with 
$$
\rho_{gg}\left(t\right) = \mathcal{N}^{2}\sum_{n=0}^{\infty}\left|\gamma_{n}\right|^{2}\left(\left|A_{23}^{\left(n\right)}\right|^{2}
+\left|A_{24}^{\left(n\right)}\right|^{2}\right),
$$
$$
\rho_{ee}\left(t\right) = \mathcal{N}^{2}\left(\sum_{n=0}^{\infty}\left|\gamma_{n+1}\right|^{2}\left|A_{12}^{\left(n+1\right)}\right|^{2}
+\sum_{n=0}^{\infty}\left|\gamma_{n}\right|^{2}\left|A_{22}^{\left(n\right)}\right|^{2}\right),
$$
and
$$
\rho_{eg}\left(t\right)=\mathcal{N}^{2}\sum_{n=0}^{\infty}\gamma_{n+1}\gamma_{n}^{*}\left(A_{12}^{\left(n+1\right)}A_{23}^{\left(n\right)*}
+A_{22}^{\left(n+1\right)}A_{24}^{\left(n\right)*}\right).
$$

\section*{Acknowledgments}

The authors would like to thank CNPq (Conselho Nacional de 
Desenvolvimento Cient\'\i fico e Tecnol\'ogico), Brazil,
for financial support through the National Institute for Science and 
Technology of Quantum Information (INCT-IQ under grant 465469/2014-0).

\bibliographystyle{elsarticle-num} 
\bibliography{refsspse}

\begin{thebibliography}{10}
\expandafter\ifx\csname url\endcsname\relax
  \def\url#1{\texttt{#1}}\fi
\expandafter\ifx\csname urlprefix\endcsname\relax\def\urlprefix{URL }\fi
\expandafter\ifx\csname href\endcsname\relax
  \def\href#1#2{#2} \def\path#1{#1}\fi

\bibitem{barnett17}
S.~M. Barnett, A.~Beige, A.~Ekert, B.~M. Garraway, C.~H. Keitel, V.~Kendon,
  M.~Lein, G.~J. Milburn, H.~M. Moya-Cessa, M.~Murao, J.~K. Pachos, G.~M.
  Palma, E.~Paspalakis, S.~J. Phoenix, B.~Piraux, M.~B. Plenio, B.~C. Sanders,
  J.~Twamley, A.~Vidiella-Barranco, M.~Kim,
  \href{https://www.sciencedirect.com/science/article/pii/S0079672717300277}{Journeys
  from quantum optics to quantum technology}, Progress in Quantum Electronics
  54 (2017) 19--45.
\newblock \href
  {https://doi.org/https://doi.org/10.1016/j.pquantelec.2017.07.002}
  {\path{doi:https://doi.org/10.1016/j.pquantelec.2017.07.002}}.
\newline\urlprefix\url{https://www.sciencedirect.com/science/article/pii/S0079672717300277}

\bibitem{giulini03}
E.~Joos, H.~D. Zeh, C.~Kiefer, D.~Giulini, J.~Kupsch, I.-O. Stamatescu,
  \href{https://link.springer.com/book/10.1007/978-3-662-05328-7}{Decoherence
  and the Appearance of a Classical World in Quantum Theory}, Springer Berlin,
  Heidelberg, 2003.
\newblock \href {https://doi.org/https://doi.org/10.1007/978-3-662-05328-7}
  {\path{doi:https://doi.org/10.1007/978-3-662-05328-7}}.
\newline\urlprefix\url{https://link.springer.com/book/10.1007/978-3-662-05328-7}

\bibitem{eisert18}
P.~Boes, H.~Wilming, R.~Gallego, J.~Eisert,
  \href{https://link.aps.org/doi/10.1103/PhysRevX.8.041016}{Catalytic quantum
  randomness}, Phys. Rev. X 8 (2018) 041016.
\newblock \href {https://doi.org/10.1103/PhysRevX.8.041016}
  {\path{doi:10.1103/PhysRevX.8.041016}}.
\newline\urlprefix\url{https://link.aps.org/doi/10.1103/PhysRevX.8.041016}

\bibitem{dorner07}
D.~Akoury, K.~Kreidi, T.~Jahnke, T.~Weber, A.~Staudte, M.~Schöffler,
  N.~Neumann, J.~Titze, L.~P.~H. Schmidt, A.~Czasch, O.~Jagutzki, R.~A.~C.
  Fraga, R.~E. Grisenti, R.~D. Muiño, N.~A. Cherepkov, S.~K. Semenov,
  P.~Ranitovic, C.~L. Cocke, T.~Osipov, H.~Adaniya, J.~C. Thompson, M.~H.
  Prior, A.~Belkacem, A.~L. Landers, H.~Schmidt-Böcking, R.~Dörner,
  \href{https://www.science.org/doi/abs/10.1126/science.1144959}{The simplest
  double slit: Interference and entanglement in double photoionization of
  $\mbox{H}_2$}, Science 318~(5852) (2007) 949--952.
\newblock \href {https://doi.org/10.1126/science.1144959}
  {\path{doi:10.1126/science.1144959}}.
\newline\urlprefix\url{https://www.science.org/doi/abs/10.1126/science.1144959}

\bibitem{roversi03}
J.~Roversi, A.~Vidiella-Barranco, H.~Moya-Cessa,
  \href{https://www.worldscientific.com/doi/10.1142/S0217984903005147}{Dynamics
  of two atoms coupled to a cavity field}, Modern Physics Letters B 17~(5-6)
  (2003) 219--224.
\newblock \href {https://doi.org/10.1142/s0217984903005147}
  {\path{doi:10.1142/s0217984903005147}}.
\newline\urlprefix\url{https://www.worldscientific.com/doi/10.1142/S0217984903005147}

\bibitem{palma08}
G.~Gennaro, G.~Benenti, G.~M. Palma,
  \href{https://doi.org/10.1209/0295-5075/82/20006}{Entanglement dynamics and
  relaxation in a few-qubit system interacting with random collisions}, {EPL}
  (Europhysics Letters) 82~(2) (2008) 20006.
\newblock \href {https://doi.org/10.1209/0295-5075/82/20006}
  {\path{doi:10.1209/0295-5075/82/20006}}.
\newline\urlprefix\url{https://doi.org/10.1209/0295-5075/82/20006}

\bibitem{lombardi10}
M.~Castagnino, S.~Fortin, O.~Lombardi,
  \href{https://www.worldscientific.com/doi/10.1142/S0217732310032664}{Is the
  decoherence of a system the result of its interaction with the environment?},
  Modern Physics Letters A 25~(17) (2010) 1431--1439.
\newblock \href {https://doi.org/10.1142/S0217732310032664}
  {\path{doi:10.1142/S0217732310032664}}.
\newline\urlprefix\url{https://www.worldscientific.com/doi/10.1142/S0217732310032664}

\bibitem{vidiella14}
A.~Vidiella-Barranco,
  \href{https://www.sciencedirect.com/science/article/pii/S0378437114000958}{Deviations
  from reversible dynamics in a qubit–oscillator system coupled to a very
  small environment}, Physica A: Statistical Mechanics and its Applications 402
  (2014) 209--215.
\newblock \href {https://doi.org/https://doi.org/10.1016/j.physa.2014.02.004}
  {\path{doi:https://doi.org/10.1016/j.physa.2014.02.004}}.
\newline\urlprefix\url{https://www.sciencedirect.com/science/article/pii/S0378437114000958}

\bibitem{decordi18}
G.~L. De{\c{c}}ordi, A.~Vidiella-Barranco,
  \href{https://doi.org/10.1080/09500340.2018.1471172}{A simple model for a
  minimal environment: the two-atom tavis{\textendash}cummings model
  revisited}, Journal of Modern Optics 65~(16) (2018) 1879--1889.
\newblock \href {https://doi.org/10.1080/09500340.2018.1471172}
  {\path{doi:10.1080/09500340.2018.1471172}}.
\newline\urlprefix\url{https://doi.org/10.1080/09500340.2018.1471172}

\bibitem{mirkin21}
N.~Mirkin, D.~Wisniacki,
  \href{https://link.aps.org/doi/10.1103/PhysRevE.103.L020201}{Quantum chaos,
  equilibration, and control in extremely short spin chains}, Phys. Rev. E 103
  (2021) L020201.
\newblock \href {https://doi.org/10.1103/PhysRevE.103.L020201}
  {\path{doi:10.1103/PhysRevE.103.L020201}}.
\newline\urlprefix\url{https://link.aps.org/doi/10.1103/PhysRevE.103.L020201}

\bibitem{aguiar05}
L.~S. Aguiar, P.~P. Munhoz, A.~Vidiella-Barranco, J.~A. Roversi,
  \href{https://doi.org/10.1088/1464-4266/7/12/049}{The entanglement of two
  dipole{\textendash}dipole coupled atoms in a cavity interacting with a
  thermal field}, Journal of Optics B: Quantum and Semiclassical Optics 7~(12)
  (2005) S769--S771.
\newblock \href {https://doi.org/10.1088/1464-4266/7/12/049}
  {\path{doi:10.1088/1464-4266/7/12/049}}.
\newline\urlprefix\url{https://doi.org/10.1088/1464-4266/7/12/049}

\bibitem{talkner09}
G.-L. Ingold, P.~H\"anggi, P.~Talkner,
  \href{https://link.aps.org/doi/10.1103/PhysRevE.79.061105}{Specific heat
  anomalies of open quantum systems}, Phys. Rev. E 79 (2009) 061105.
\newblock \href {https://doi.org/10.1103/PhysRevE.79.061105}
  {\path{doi:10.1103/PhysRevE.79.061105}}.
\newline\urlprefix\url{https://link.aps.org/doi/10.1103/PhysRevE.79.061105}

\bibitem{ashhab14}
S.~Ashhab,
  \href{https://link.aps.org/doi/10.1103/PhysRevA.90.062120}{Landau-zener
  transitions in a two-level system coupled to a finite-temperature harmonic
  oscillator}, Phys. Rev. A 90 (2014) 062120.
\newblock \href {https://doi.org/10.1103/PhysRevA.90.062120}
  {\path{doi:10.1103/PhysRevA.90.062120}}.
\newline\urlprefix\url{https://link.aps.org/doi/10.1103/PhysRevA.90.062120}

\bibitem{vidiella16}
A.~Vidiella-Barranco,
  \href{https://www.sciencedirect.com/science/article/pii/S0378437116301649}{Evolution
  of a quantum harmonic oscillator coupled to a minimal thermal environment},
  Physica A: Statistical Mechanics and its Applications 459 (2016) 78--85.
\newblock \href {https://doi.org/https://doi.org/10.1016/j.physa.2016.04.033}
  {\path{doi:https://doi.org/10.1016/j.physa.2016.04.033}}.
\newline\urlprefix\url{https://www.sciencedirect.com/science/article/pii/S0378437116301649}

\bibitem{decordi20}
G.~De{\c{c}}ordi, A.~Vidiella-Barranco,
  \href{https://doi.org/10.1016/j.optcom.2020.126233}{Sudden death of
  entanglement induced by a minimal thermal environment}, Optics Communications
  475 (2020) 126233.
\newblock \href {https://doi.org/10.1016/j.optcom.2020.126233}
  {\path{doi:10.1016/j.optcom.2020.126233}}.
\newline\urlprefix\url{https://doi.org/10.1016/j.optcom.2020.126233}

\bibitem{ashhab06a}
S.~Ashhab, J.~R. Johansson, F.~Nori,
  \href{https://dx.doi.org/10.1088/1367-2630/8/6/103}{Rabi oscillations in a
  qubit coupled to a quantum two-level system}, New Journal of Physics 8~(6)
  (2006) 103.
\newblock \href {https://doi.org/10.1088/1367-2630/8/6/103}
  {\path{doi:10.1088/1367-2630/8/6/103}}.
\newline\urlprefix\url{https://dx.doi.org/10.1088/1367-2630/8/6/103}

\bibitem{ashhab06b}
S.~Ashhab, J.~Johansson, F.~Nori,
  \href{https://www.sciencedirect.com/science/article/pii/S0921453406005776}{Decoherence
  dynamics of a qubit coupled to a quantum two-level system}, Physica C:
  Superconductivity and its Applications 444~(1) (2006) 45--52.
\newblock \href {https://doi.org/https://doi.org/10.1016/j.physc.2006.04.106}
  {\path{doi:https://doi.org/10.1016/j.physc.2006.04.106}}.
\newline\urlprefix\url{https://www.sciencedirect.com/science/article/pii/S0921453406005776}

\bibitem{decordi17}
G.~De{\c{c}}ordi, A.~Vidiella-Barranco,
  \href{https://doi.org/10.1016/j.optcom.2016.10.017}{Two coupled qubits
  interacting with a thermal bath: A comparative study of different models},
  Optics Communications 387 (2017) 366--376.
\newblock \href {https://doi.org/10.1016/j.optcom.2016.10.017}
  {\path{doi:10.1016/j.optcom.2016.10.017}}.
\newline\urlprefix\url{https://doi.org/10.1016/j.optcom.2016.10.017}

\bibitem{horodecki01}
K.~\ifmmode~\dot{Z}\else \.{Z}\fi{}yczkowski, P.~Horodecki, M.~Horodecki,
  R.~Horodecki,
  \href{https://link.aps.org/doi/10.1103/PhysRevA.65.012101}{Dynamics of
  quantum entanglement}, Phys. Rev. A 65 (2001) 012101.
\newblock \href {https://doi.org/10.1103/PhysRevA.65.012101}
  {\path{doi:10.1103/PhysRevA.65.012101}}.
\newline\urlprefix\url{https://link.aps.org/doi/10.1103/PhysRevA.65.012101}

\bibitem{eberly04}
T.~Yu, J.~H. Eberly,
  \href{https://link.aps.org/doi/10.1103/PhysRevLett.93.140404}{Finite-time
  disentanglement via spontaneous emission}, Phys. Rev. Lett. 93 (2004) 140404.
\newblock \href {https://doi.org/10.1103/PhysRevLett.93.140404}
  {\path{doi:10.1103/PhysRevLett.93.140404}}.
\newline\urlprefix\url{https://link.aps.org/doi/10.1103/PhysRevLett.93.140404}

\bibitem{dodonov74}
V.~Dodonov, I.~Malkin, V.~Man'ko,
  \href{https://www.sciencedirect.com/science/article/pii/0031891474902158}{Even
  and odd coherent states and excitations of a singular oscillator}, Physica
  72~(3) (1974) 597--615.
\newblock \href {https://doi.org/https://doi.org/10.1016/0031-8914(74)90215-8}
  {\path{doi:https://doi.org/10.1016/0031-8914(74)90215-8}}.
\newline\urlprefix\url{https://www.sciencedirect.com/science/article/pii/0031891474902158}

\bibitem{vidiella92}
V.~Bu\ifmmode~\check{z}\else \v{z}\fi{}ek, A.~Vidiella-Barranco, P.~L. Knight,
  \href{https://link.aps.org/doi/10.1103/PhysRevA.45.6570}{Superpositions of
  coherent states: Squeezing and dissipation}, Phys. Rev. A 45 (1992)
  6570--6585.
\newblock \href {https://doi.org/10.1103/PhysRevA.45.6570}
  {\path{doi:10.1103/PhysRevA.45.6570}}.
\newline\urlprefix\url{https://link.aps.org/doi/10.1103/PhysRevA.45.6570}

\bibitem{domokos94}
P.~Domokos, J.~Janszky, P.~Adam, T.~Larsen,
  \href{https://doi.org/10.1088/0954-8998/6/3/005}{Role of quantum interference
  in producing non-classical states}, Quantum Optics: Journal of the European
  Optical Society Part B 6~(3) (1994) 187--199.
\newblock \href {https://doi.org/10.1088/0954-8998/6/3/005}
  {\path{doi:10.1088/0954-8998/6/3/005}}.
\newline\urlprefix\url{https://doi.org/10.1088/0954-8998/6/3/005}

\bibitem{hanggi10}
J.~Dajka, M.~Mierzejewski, J.~{\L}uczka, P.~Hänggi,
  \href{https://doi.org/10.1016/j.physe.2009.06.080}{Dephasing of qubits by the
  {S}chrödinger cat}, Physica E: Low-dimensional Systems and Nanostructures
  42~(3) (2010) 374--377.
\newblock \href {https://doi.org/10.1016/j.physe.2009.06.080}
  {\path{doi:10.1016/j.physe.2009.06.080}}.
\newline\urlprefix\url{https://doi.org/10.1016/j.physe.2009.06.080}

\bibitem{almeida10}
J.~S. Sales, L.~F. da~Silva, N.~G. de~Almeida,
  \href{https://dx.doi.org/10.1088/0953-4075/43/24/245504}{Preparing
  superposition of squeezed coherent states under thermal reservoir}, Journal
  of Physics B: Atomic, Molecular and Optical Physics 43~(24) (2010) 245504.
\newblock \href {https://doi.org/10.1088/0953-4075/43/24/245504}
  {\path{doi:10.1088/0953-4075/43/24/245504}}.
\newline\urlprefix\url{https://dx.doi.org/10.1088/0953-4075/43/24/245504}

\bibitem{mandel79}
L.~Mandel,
  \href{http://www.osapublishing.org/ol/abstract.cfm?URI=ol-4-7-205}{Sub-{P}oissonian
  photon statistics in resonance fluorescence}, Opt. Lett. 4~(7) (1979)
  205--207.
\newblock \href {https://doi.org/10.1364/OL.4.000205}
  {\path{doi:10.1364/OL.4.000205}}.
\newline\urlprefix\url{http://www.osapublishing.org/ol/abstract.cfm?URI=ol-4-7-205}

\bibitem{horodecki09}
R.~Horodecki, P.~Horodecki, M.~Horodecki, K.~Horodecki,
  \href{https://link.aps.org/doi/10.1103/RevModPhys.81.865}{Quantum
  entanglement}, Rev. Mod. Phys. 81 (2009) 865--942.
\newblock \href {https://doi.org/10.1103/RevModPhys.81.865}
  {\path{doi:10.1103/RevModPhys.81.865}}.
\newline\urlprefix\url{https://link.aps.org/doi/10.1103/RevModPhys.81.865}

\bibitem{wootters97}
S.~A. Hill, W.~K. Wootters,
  \href{https://link.aps.org/doi/10.1103/PhysRevLett.78.5022}{Entanglement of a
  pair of quantum bits}, Phys. Rev. Lett. 78 (1997) 5022--5025.
\newblock \href {https://doi.org/10.1103/PhysRevLett.78.5022}
  {\path{doi:10.1103/PhysRevLett.78.5022}}.
\newline\urlprefix\url{https://link.aps.org/doi/10.1103/PhysRevLett.78.5022}

\bibitem{plenio14}
T.~Baumgratz, M.~Cramer, M.~Plenio,
  \href{https://doi.org/10.1103/physrevlett.113.140401}{Quantifying coherence},
  Physical Review Letters 113~(14) (sep 2014).
\newblock \href {https://doi.org/10.1103/physrevlett.113.140401}
  {\path{doi:10.1103/physrevlett.113.140401}}.
\newline\urlprefix\url{https://doi.org/10.1103/physrevlett.113.140401}

\bibitem{plenio17}
A.~Streltsov, G.~Adesso, M.~B. Plenio,
  \href{https://link.aps.org/doi/10.1103/RevModPhys.89.041003}{Colloquium:
  Quantum coherence as a resource}, Rev. Mod. Phys. 89 (2017) 041003.
\newblock \href {https://doi.org/10.1103/RevModPhys.89.041003}
  {\path{doi:10.1103/RevModPhys.89.041003}}.
\newline\urlprefix\url{https://link.aps.org/doi/10.1103/RevModPhys.89.041003}

\end{thebibliography}





\end{document}